\documentclass[12pt, draftclsnofoot, onecolumn]{IEEEtran} 
\usepackage{setspace}
\usepackage{multirow}
\usepackage{bm}
\usepackage{amsmath}

\usepackage{amsmath, cite}
\usepackage{algpseudocode}
\usepackage{graphicx}
\usepackage{subcaption}
\usepackage{enumitem}

\usepackage{array}
\usepackage{amssymb}
\usepackage{textcomp}
\usepackage{color}
\usepackage{epstopdf}
\usepackage[justification=centering]{caption}
\usepackage{bbm}
\usepackage{tabularx}
\usepackage{tablefootnote}
\usepackage{stfloats}
\usepackage{multirow}
\usepackage[table]{xcolor}
\usepackage{hhline}
\usepackage{booktabs}
\usepackage{subcaption}
\usepackage{cleveref}

\makeatletter

\newcommand{\Rmnum}[1]{\expandafter\@slowromancap\romannumeral #1@}

\makeatother

\begin{document}
	
	\title{Low Power Receiver Front Ends: Scaling Laws and Applications}
	
	
	\author{\IEEEauthorblockN{Muris Sarajli\'{c}, \textit{Student Member, IEEE}, Liang Liu, \textit{Member, IEEE}, Henrik Sj\"{o}land, \textit{Senior Member, IEEE} and Ove Edfors, \textit{Senior Member, IEEE}} \\		
		\thanks{This work was supported by the Swedish Foundation for Strategic Research (SSF), in the scope of the Digitally Assisted Radio Evolution (DARE) project.
		
		Corresponding author is Muris Sarajli\'{c} (e-mail: muris.sarajlic@eit.lth.se).
		
		M. Sarajli\'{c}, L. Liu and O. Edfors are with the Department of Electrical and Information Technology, Lund University, SE-221 00 Lund, Sweden.
	
		H. Sj\"{o}land is with Ericsson, SE-221 83 Lund, Sweden and with the Department of Electrical and Information Technology, Lund University, SE-221 00 Lund, Sweden.}}	
	\maketitle
	
	
	\begin{abstract}
		In this paper, we combine communication-theoretic laws with known, practically verified results from circuit theory.								
		As a result, we obtain closed-form theoretical expressions linking fundamental system design and environment parameters with the power consumption of analog front ends for communication receivers.			
		This collection of scaling laws and bounds is meant to serve as a theoretical reference for practical low power front end design.		
		In one set of results, we first find that the front end power consumption scales at least as $\mathit{SNDR}^{3/2}$ if environment parameters (fading and blocker levels) are static.			
		The obtained scaling law is subsequently used to derive relations between front end power consumption and several other important communication system parameters, namely, digital modulation constellation size, symbol error probability, error control coding gain and coding rate.	
		Such relations, in turn, can be used when deciding which system design strategies to adopt for low-power applications.			
		For example, if error control coding is employed, the most energy-efficient strategy for the entire receiver is to use codes with moderate coding gain and simple decoding algorithms, such as convolutional codes.		
		In another collection of results, we find how front end power scales with environment parameters if the performance is kept constant.
		This yields bounds on average power reduction of receivers that adapt to the communication environment.	
		For instance, if a receiver front end adapts to fading fluctuations while keeping the performance above some given minimum requirement, power can theoretically be reduced  at least 20x compared to a non-adaptive front end.
	\end{abstract}	
	
	
	\IEEEpeerreviewmaketitle
	
	\section{Introduction}
	
	\IEEEPARstart{L}{ow} power consumption is one of the main design targets for communication receivers, and its importance is especially high when it comes to wireless devices, which are often battery-powered and therefore energy-limited.
	At the same time, receivers also need to satisfy some performance requirement, such as minimum throughput and maximum bit error rate (BER).
	
	Designing receivers that jointly meet power consumption and performance criteria tends to be predominantly based on the experience of hardware designers.
	Additionally, more often than not, receiver designs are optimized based on the worst-case scenario of operation (the most adverse possible combination of environment conditions under which satisfying performance must be delivered).
	The latter, conservative design trend in particular is what prevents hardware designs from exploiting their full potential for low-power operation.
	
	It would be of significant interest to be able to theoretically predict how much power would be consumed by a receiver with certain performance requirements, with all system and environment constraints taken into consideration.
	Such a result would serve as a benchmark and motivation for both practical hardware and system design, an indicator of how low the power consumption can really be driven.	
	If combined with the knowledge of the statistical properties of environment variables, it could also provide a measure of how much power can be saved if the receiver adapts to the communication environment.
		
	The analog front end (AFE) (the chain of analog signal blocks of the receiver excluding the oscillator) typically has a defining impact on the overall performance of the receiver, while also consuming a substantial portion of its power.
	One of the main questions of low-power receiver design can thus be formulated as
	\begin{center}
		\textit{``How does the power consumption of an analog front end (AFE) of a receiver scale with performance?''}
	\end{center}
	
	If this question is answered, some important follow-up questions can be answered as well, such as:
	
	\begin{itemize}
		\item If the overall system design features techniques that serve to improve performance (e.g. by use of error control coding) and this opens up the possibility of relaxing the design of the AFE, how much power do we save by performing this relaxation?
		\item If the receiver is made adaptive to communication environment conditions (e.g. channel fading or out-of-band interference), how much AFE power can be saved, on average, compared to a design based on worst-case conditions?
		
	\end{itemize}
	
	The theoretical analysis of the relation between analog circuit power consumption and performance appears not to have gained a lot of attention in the scientific community.
	The relation between power consumption and performance for individual analog blocks is analyzed in \cite{Abidi2000} and \cite{Svensson2015}.
	It is found that power consumption grows linearly with the dynamic range of an analog circuit block\footnote{The definitions of dynamic range differ slightly between these two papers. As will be shown, here we adhere to the definition given in \cite{Abidi2000}.}.
	Analysis of this relation for a chain of analog blocks becomes rather involved because performance metrics for the entire chain exhibit a complex dependence on gain, noise and linearity properties for individual blocks.
	Moreover, there are practically infinitely many combinations of per-block parameters that satisfy the performance requirements for the entire chain, with each combination resulting in a unique power consumption.
	A sensible approach is then to find the combination that yields the minimum power consumption, which then makes it possible to reveal the implicit or explicit connection between the performance requirement and the obtained optimal power consumption.
	In \cite{Sheng2006} and \cite{Baltus2000} this approach is adopted, with the focus being mostly on how to conveniently model the power - performance relation for individual blocks and how to solve the optimization problem.	
	The analysis in \cite{Heuvel2014} extends the ideas from \cite{Baltus2000}, with the power - performance relation also being given some treatment in the context of communication systems.	
	
	There also exists a body of academic work \cite{Meghdadi2014} - \cite{Banerjee2017} that examines the topic of environment-adaptive AFEs and receivers, with the focus being primarily on practical hardware implementations.
	It is demonstrated that adaptive receivers are implementable, and various implementation strategies are suggested.
	Furthermore, measured power numbers from these designs indicate that substantial power reduction is attainable if environment-adaptive receiver techniques are adopted.	
	
	What is found to be largely missing in the existing literature is a work that takes the power-performance laws from circuit theory and combines them with classical results from communication theory to formulate joint circuit-communication-theoretical laws of system behavior \footnote{A rare example is \cite{Heuvel2014}, but with the analysis limited only to the connection between throughput and relative level of the out-of-band blocking signal.}.
	With such laws at hand, system design questions such as ``if the BER requirement is relaxed from $10^{-6}$ to $10^{-3}$ and we redesign the receiver to meet the new requirements, how much power is this new receiver expected to consume?'' can be answered in a precise and immediate fashion, without resorting to educated guessing or iterative hardware redesign and simulation/measurement cycles.
	Moreover, by taking into account the influence of environment conditions on front end power consumption, it would be possible to precisely determine power savings obtainable when making the receiver environment-adaptive.
	
	Here we aim at bridging this gap between circuit and communication theory.	
	The idea is to obtain theoretical expressions that will describe how optimal AFE power consumption scales with important system and environment parameters.			
	More specifically, we are interested in finding out the scaling of front end power with the signal-to-noise-and-distortion ratio (SNDR), representing system performance, when the environment parameters (fading and out-of-band interference) do not exhibit temporal changes.	
	Conversely, we also aim to describe how AFE power scales with environment parameters when SNDR is kept constant.
	The obtained set of fundamental scaling laws can then be used to build up a more extensive system level analysis.  
	We derive our scaling laws from a known relation between performance and minimum power consumption for AFEs, presented and verified in actual hardware implementations in \cite{Sheng2006}.	
	This relation is modified so that it can be seamlessly combined with communication-theoretic laws.		
	One set of results is based on a novel scaling law we obtain, namely, that AFE power consumption scales at least as $\mathit{SNDR}^{3/2}$.	
	This result is then employed in finding closed-form expressions for AFE power scaling with QAM constellation size, symbol error rate and error control coding gain, which are further used to decide on appropriate system-level strategies for low-power design.		
	In another line of results, we obtain power-law type relations between AFE power and environment parameters.	
	These are combined with fading and blocker statistics, yielding important theoretical bounds on average power savings of environment-adaptive front ends, which demonstrate that substantial power savings are possible if the environment-adaptive design approach is adopted.	
	
	Throughout the course of our analysis, we rely on the fact that the fundamental results we build upon have been verified in practical front end implementations and we do not aim at recreating these verifications.		
	Instead, we put the focus on laying out a general theoretical framework for low power receiver design and showing the advantages of environment-adaptive designs, which will hopefully make this work both a point of reference and motivation for future research efforts in the area of practical hardware implementations of such systems.	
	
	
	\section{Optimal power consumption of analog front ends}
	\label{sec:section2}
	
	Let us observe a chain of analog circuit blocks that form the front end of a communications receiver.
	One example of such a chain can be the direct conversion receiver with the structure LNA - downconversion mixer - channel select filter - variable gain amplifier.
	While the direct conversion receiver is given as an example, we emphasize that the forthcoming analysis holds for any type of receiver chain.
	
	Each of the blocks in the chain can be qualitatively characterized by noise and linearity properties (serving as performance quantifiers) and by an associated power consumption.
	Noise performance is commonly quantified by noise power spectral density $\overline{V}_{\textnormal{N}}^2 ~ \left [ {\textnormal{V}^2}/{\textnormal{Hz}} \right ]$, and
	linearity by $V_{\textnormal{IIP3}}^2 ~ [\textnormal{V}^2]$, the input-referred third-order intercept voltage squared.		
	Additionally, we denote by $F_{\textnormal{AFE}}$ the total noise factor and by $V_{\textnormal{IIP3, AFE}}^2$ the total IIP3 voltage squared of the AFE chain.
	These are usually set by performance requirements dictated from digital baseband.		
	Given $F_{\textnormal{AFE}}$ and $V_{\textnormal{IIP3, AFE}}^2$, one would preferably like to select $\overline{V}_{\textnormal{N}}^2$ and $V_{\textnormal{IIP3}}^2$ of individual blocks such that the power consumption of the entire chain is minimized.
	
	In order to solve this task, we first need to look into the nature of the relation between the performance quantifiers and power consumption for each block.	
	The dynamic range of a block with index $j$ is defined as
	\begin{equation}
	\label{eq:dynamic_range}
	\mathit{DR}_j \triangleq \frac{V_{\textnormal{IIP3}, j}^2}{\overline{V}_{\textnormal{N}, j}^2}.
	\end{equation}
	As presented in \cite{Abidi2000} and \cite{Sheng2006}, for a wide range of the most common front-end blocks, the power consumption of a circuit is linear with the dynamic range as defined in (\ref{eq:dynamic_range}), i.e.	
	\begin{equation}
	\label{eq:DR_and_P}
	P_j = P_{\textnormal{C},j}^{} \mathit{DR}_j,
	\end{equation}
	where $P_{\textnormal{C}, j}^{}$ is a proportionality factor that can be taken as a natural figure-of-merit for analog blocks.
	
	Starting from this simple but powerful relation, the authors in \cite{Sheng2006} have devised a method of finding $\overline{V}_{\textnormal{N}, j}^2$ and $V_{\textnormal{IIP3}, j}^2$ that results in minimum power consumption of the whole AFE chain.		
	Although proof is given in \cite{Sheng2006} that relation (\ref{eq:DR_and_P}) holds for standard CMOS circuits (such as a common-source stage LNA, a double-balanced Gilbert cell mixer and an OTA-C baseband filter), the results of the optimization are valid for any chain of analog blocks that satisfy (\ref{eq:DR_and_P}) and are hence not limited only to CMOS circuits.	
	Moreover, \cite{Sheng2006} provides a comparison of theoretically optimal $\overline{V}^2_{\textnormal{N}, j}$ and $V_{\textnormal{IIP3}, j}^2$ with measured noise PSD and IIP3 from an actual ``hand-optimized'' Bluetooth receiver implementation, with a good match between the two.
	This hardware verification naturally extends to our analysis, which considers optimally designed front ends in communication system settings.	
 		
	What is important for our analysis is that the method from \cite{Sheng2006} provides the connection between the optimal power consumption of the entire AFE, denoted by $P_{\textnormal{AFE}}^*$, and $V_{\textnormal{IIP3, AFE}}^2$ and $F_{\textnormal{AFE}}$, which reads \cite[eq. (60)]{Sheng2006}
	\begin{equation}
	\label{eq:total_P}
	P_{\textnormal{AFE}}^* = \frac {V_{\textnormal{IIP3, AFE}}^2}{(F_{\textnormal{AFE}} - 1)kT50} \left ( \sum_{j=1}^N \sqrt[3]{P_{\textnormal{C}, j}}  \right )^3,
	\end{equation}
	where $k$ is Boltzmann constant and $T$ temperature in Kelvins.
	Remarkably, the optimal power consumption of the chain is independent of power/voltage gains of individual blocks.
	
	If we are to use the result in (\ref{eq:total_P}) for drawing conclusions on the system-level behaviour of receivers, it would be convenient to ``translate'' this result to system designer parlance, so that it features power-related parameters:
	\begin{itemize}
		\item received wanted signal power at the antenna - $p_\textnormal{S}^{}$,
		\item total input-referred thermal noise power - $p_\textnormal{N}^{}$,
		\item power of the out-of-band (OOB) interfering signal at the antenna - $p_\textnormal{I}^{}$.
		\footnote{The results (\ref{eq:dynamic_range}) and consequently, (\ref{eq:total_P}) were derived with the assumption of a two-tone interference model.
			For the sake of consistency, we maintain this model throughout our analysis, and $p_\textnormal{I}^{}$ then denotes the total power of the two interfering tones.
			However, we conjecture that the obtained trends hold even in the case of modulated interferers.}
	\end{itemize}	
	As a first step, we can relate $p_\textnormal{N}^{}$ and $F_{\textnormal{AFE}}^{}$ through
	\begin{equation}
	\label{eq:noise_power_F_connection}
		p_{\textnormal{N}}^{} = k T B F_\textnormal{AFE}^{},
	\end{equation}
	with $B$ being the noise-equivalent bandwidth of the system.
	On the other hand, IIP3 power and voltage can be related by
	\begin{equation}
	\label{eq:IIP3_voltage_power_connection}
	p_{\textnormal{IIP3, AFE}}^{} = \frac {V_{\textnormal{IIP3, AFE}}^2 }{R_{\textnormal{in}}},
	\end{equation}
	where $R_{\textnormal{in}}$ is the input resistance of the receiver which we assume to be 50 $\Omega$ for simplicity.
	In order to directly assess the impact of third-order nonlinearity on system performance, we need to relate the IIP3 to $p_{\textnormal{IM3}}^{}$, the power of the in-band third-order intermodulation (IM3) distortion.
	A well-known relation linking $p_{\textnormal{IIP3}}^{}$, $p_\textnormal{I}^{}$ and  $p_{\textnormal{IM3}}^{}$ reads \cite{Razavi2011}
	\begin{equation}
	\label{eq:IIP_pi_connection}
	p_{\textnormal{IIP3}}^{} = \sqrt {\frac {p_\textnormal{I}^3}{p_{\textnormal{IM3}}^{}}}.
	\end{equation}	
	For the purpose of notational convenience, we denote the last term in (\ref{eq:total_P}) as
	\begin{equation}
	\kappa_{\textnormal{circuit}}^{} \triangleq \left ( \sum_{j=1}^N \sqrt[3]{P_{\textnormal{C}, j}^{}}  \right )^3
	\end{equation}
	and use \eqref{eq:noise_power_F_connection}, \eqref{eq:IIP3_voltage_power_connection} and \eqref{eq:IIP_pi_connection} in conjunction with \eqref{eq:total_P} to obtain
	\begin{equation}
	\label{eq:P_AFE_final_1}
	P_{\textnormal{AFE}}^* = \frac {F_{\textnormal{AFE}}^{}}{F_{\textnormal{AFE}^{}} - 1} B \frac {p_\textnormal{I}^{3/2}}{p_\textnormal{N}^{} \sqrt{p_{\textnormal{IM3}}}^{}} \kappa_{\textnormal{circuit}}^{}.
	\end{equation}	
	For the analysis at hand it is of use to define the power ratio of intermodulation distortion and noise
	\begin{equation}
	 \label{eq:alpha_def}
	 \alpha_{\textnormal{IM3}}^{} \triangleq \frac {p_{\textnormal{IM3}}^{}}{p_\textnormal{N}^{}},
	\end{equation}
	which combined with \eqref{eq:P_AFE_final_1} yields	
	\begin{equation}
	\label{eq:final_form_1}
	P_{\textnormal{AFE}}^{*} = \frac {F_{\textnormal{AFE}}^{}}{F_{\textnormal{AFE}^{}}-1} B \frac {1}{\sqrt{\alpha_{\textnormal{IM3}}^{}}} \left ( \frac {p_\textnormal{I}^{}}{p_\textnormal{N}^{}}  \right )^{3/2} \kappa_{\textnormal{circuit}}^{},
	\end{equation}
	with $p_\textnormal{I}^{} > 0$ which follows from constraint $V_{\textnormal{IIP3, AFE}}^2 > 0$.	
	Equation (\ref{eq:final_form_1}) can be used as a basis for deriving simple but very useful scaling laws, as presented in the following section.

	\section{Scaling laws of AFE power consumption}
	\label{sec:scaling_laws}
	
	A holistic receiver system design benefits greatly from the availability of closed form relations between receiver power consumption and other system parameters.
	This way, a mathematically tractable analysis of the tradeoffs encountered during receiver system design is made possible.	
	When it comes to real-world hardware, obtaining such relations is not a trivial task, and there always exists a tradeoff between the accuracy of the functional dependencies and their analytical tractability.
	Ideally, they should appear in form of simple power laws.
	It turns out that (\ref{eq:final_form_1}), under some realistic assumptions, can yield such simple relations.
	The advantage of using (\ref{eq:final_form_1}) for this purpose is that it is soundly grounded in circuit theory which has also been verified against real-life receiver designs, so it enables striking a good balance between accuracy, simplicity and theoretical rigour.
	
	
	\begin{figure*}[t!]
		\centering      
		\includegraphics[width=0.85\textwidth]{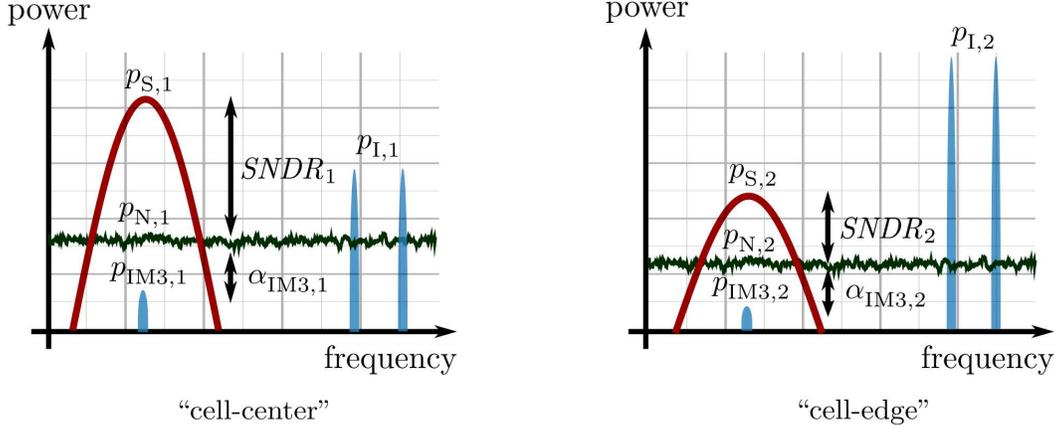}          
		\caption{Illustration of all relevant system parameters for the cell center scenario (left: strong wanted signal, weak OOB interference, high $\mathit{SNDR}$ requirement) and cell edge scenario (right: weak wanted signal, strong OOB interference, low $\mathit{SNDR}$ requirement).}
		\label{fig:two_setups_sketch}
	\end{figure*}	

	To start with, a performance metric is needed that will provide a link between baseband metrics, like bit error rate ($\mathit{BER}$), via power-related system parameters, with circuit parameters $F_{\textnormal{AFE}}^{}$ and $V_{\textnormal{IIP3, AFE}}^2$.
	A commonly used such metric is the signal-to-noise-and-distortion ratio\footnote{It is commonly assumed that the third-order distortion is the dominant nonlinear impairment in analog systems.
	Therefore, along with thermal noise, we consider it a determining factor of system performance.
	All other possible impairments, such as second-order distortion, flicker noise, phase noise--either in-band or due to reciprocal mixing--are through appropriate design assumed to be dominated by thermal noise and third-order distortion in all scenarios considered.}, which is defined as 
	\begin{equation}
	\label{eq:SNDR_definition}
		\mathit{SNDR} \triangleq \frac {p_\textnormal{S}^{}}{p_\textnormal{N}^{} + p_{\textnormal{IM3}}^{}} = \frac {p_\textnormal{S}^{}}{\left ( 1 + \alpha_{\textnormal{IM3}}^{}  \right ) p_\textnormal{N}^{}}.
	\end{equation}
	
	Now we focus our attention on four fundamental receiver design parameters, namely, $\mathit{SNDR}$ and $B$ (the values of which are determined by the particular application), and $p_\textnormal{S}^{}$ and $p_\textnormal{I}^{}$ (which describe the environment and are generally stochastic).
	The values of the fundamental parameters define distinct application-environment scenarios.
	We structure our analysis around a pair of such scenarios: an initial (pre-scaling) and target (post-scaling) scenario.
	An illustration of the relations between parameters of importance for an example scenario pair is given in Fig.~\ref{fig:two_setups_sketch}.
	For each of the two scenarios--under practical constraints on parameter values--we assume that an  analog front end with minimal power consumption is designed using the procedure described in \cite{Sheng2006}.
	Our aim is relating the scaling of fundamental parameter values between the two scenarios and the scaling of optimal front end power.
	To this end, we label variables corresponding to pre-scaling and post-scaling scenarios with indices 1 and 2, respectively.	
	The scaling of the optimal power consumption is denoted as
	\begin{equation}
		\varsigma_{\textnormal{P}}^{} \triangleq \frac {P_{\textnormal{AFE}, 2}^*}{P_{\textnormal{AFE}, 1}^*}.
	\end{equation}	
	The scaling factors of bandwidth, $\mathit{SNDR}$, signal and interference power are defined analogously to $\varsigma_{\textnormal{P}}^{}$ and denoted respectively as $\varsigma_{\textnormal{B}}^{}$, $\varsigma_{\textnormal{SNDR}}^{}$, $\varsigma_{\textnormal{S}}^{}$ and $\varsigma_{\textnormal{I}}^{}$.
	By using \eqref{eq:final_form_1} and \eqref{eq:SNDR_definition}, the scaling of front end power reads
	\begin{equation}
	\label{eq:original_power_ratio}
		\varsigma_{\textnormal{P}}^{} = \varphi~\delta~\varsigma_{\textnormal{B}}^{}  \varsigma_{\textnormal{SNDR}}^{3/2}~\varsigma_{\textnormal{I}}^{3/2}~\varsigma_{\textnormal{S}}^{-3/2},		
	\end{equation}	
	where, for analytical convenience, we have introduced the factors
	\begin{equation}
	\label{eq:phi_definition}
	\varphi \triangleq \frac {F_{\textnormal{AFE}, 2}^{}}{F_{\textnormal{AFE}, 1}^{}} \frac {F_{\textnormal{AFE}, 1}^{} - 1}{F_{\textnormal{AFE}, 2}^{} - 1}
	\end{equation}
	and
	\begin{equation}
	\label{eq:delta_def}
		\delta \triangleq \sqrt {\frac {\alpha_{\textnormal{IM3}, 1}^{}}{\alpha_{\textnormal{IM3}, 2}^{}}} \left ( \frac {1 + \alpha_{\textnormal{IM3}, 2}^{}} {1 + \alpha_{\textnormal{IM3}, 1}^{}}  \right )^{3/2}.
	\end{equation}	
	
		\begin{table*}[t]
		\centering
		\scriptsize	
		\caption{Collection of fundamental scaling laws for front end power. Application and environment constraints (columns 1-4) are translated to front end design requirements (columns 4-8). Consequently, if the front ends are designed optimally, their power scales with a selected parameter (bandwidth, $\mathit{SNDR}$, received power, blocker power) as given in columns 9 and 10.}
		\label{tab:scaling_laws}	
		\begin{tabular}{|c|c|c|c|c|c|c|c|c|c|}
			\noalign{\hrule height 2pt}
			
			\multicolumn{3}{| c }{ {\cellcolor[gray]{.75}}{\parbox[c][1cm][c]{1cm} \centering Constraints}} &
			\multicolumn{1}{ c }{ {\cellcolor[gray]{.85}}{\parbox[c][1cm][c]{0.6cm} \centering }} &
			\multicolumn{4}{ c |}{{\cellcolor[gray]{.95}}{\parbox[c][1cm][c]{2cm} \centering Design requirements}}  &
			\multicolumn{1}{ c |}{} &
			\multicolumn{1}{ c |}{}
			\\			
						
			\multicolumn{1}{|c}{{\cellcolor[gray]{.9}}{\parbox[c][1cm][c]{1.4cm}{\centering Performance}}} &
			\multicolumn{2}{c}{{\cellcolor[gray]{.90}}{\parbox[c][1.1cm][c]{1.3cm}{\centering Environment}}} &
			\multicolumn{3}{ c}{{\cellcolor[gray]{.8}} {\parbox[c][1.1cm][c]{1.8cm}{\centering System}}} &
			\multicolumn{2}{c|}{{\cellcolor[gray]{.90}}{\parbox[c][1.1cm][c]{1.6cm}{\centering Circuit}}} &
			\multirow{3}{*}{\parbox[c][1.1cm][c]{1.8cm}{\centering Power \\ scaling}} &
			\multirow{3}{*}{\parbox[c][1.1cm][c]{2cm}{\centering Properties of $\varphi$}} \\ 
			
			\parbox[c][1cm][c]{1.4cm}{\centering $\mathit{SNDR}_2^{}$} &
			{\parbox[c][1cm][c]{0.5cm}{\centering $p_{\textnormal{S}, 2}^{}$}} &
			{\parbox[c][1cm][c]{0.5cm}{\centering $p_{\textnormal{I}, 2}^{}$}} &
			{\parbox[c][1cm][c]{0.5cm}{\centering $B_2$}} &
			{\parbox[c][1cm][c]{0.6cm}{\centering $p_{\textnormal{N}, 2}^{}$}} &	
			{\parbox[c][1cm][c]{0.8cm}{\centering $p_{\textnormal{IM3}, 2}^{}$}} &			
			{\parbox[c][1cm][c]{0.6cm}{\centering $F_2$}} &	
			{\parbox[c][1cm][c]{1.5cm}{\centering $V_{\textnormal{IIP3}, 2}^2$}} & & \\ 
			\hline
			\parbox[c][1cm][c]{1.4cm}{\centering $\mathit{SNDR}_1$} &
			{\parbox[c][1cm][c]{0.5cm}{\centering $p_{\textnormal{S}, 1}^{}$}} &
			{\parbox[c][1cm][c]{0.5cm}{\centering $p_{\textnormal{I}, 1}^{}$}} &
			{\parbox[c][0.7cm][c]{0.5cm}{\centering $\varsigma_\textnormal{B}^{} B_1$}} &
			{\parbox[c][1cm][c]{0.6cm}{\centering $p_{\textnormal{N}, 1}^{}$}} &	
			{\parbox[c][1cm][c]{0.8cm}{\centering $p_{\textnormal{IM3}, 1}^{}$}} &
			{\parbox[c][1cm][c]{0.6cm}{\centering \begin{equation*} \frac {F_1} {\varsigma_{\textnormal{B}}} \end{equation*} }} &	
			$V_{\textnormal{IIP3}, 1}^2$	&			
			{\parbox[c][1cm][l]{1.8cm}{\centering \begin{equation} \label{eq:scaling_law_B} \varsigma_{\textnormal{P}}^{} = \varphi \varsigma_{\textnormal{B}}^{} \end{equation}}} &
			$\begin{array} {rl} \varphi < 1, & 0 < \varsigma_{\textnormal{B}}^{} < 1 \\ \varphi \geq 1, & 1 \leq \varsigma_{\textnormal{B}}^{} < F_1 \end{array}$ \\
			
			{\parbox[c][0.7cm][c]{1.5cm}{\centering $\varsigma_\textnormal{SNDR}^{} \mathit{SNDR}_1$}} &
			{\parbox[c][1cm][c]{0.5cm}{\centering $p_{\textnormal{S}, 1}^{}$}} &
			{\parbox[c][1cm][c]{0.5cm}{\centering $p_{\textnormal{I}, 1}^{}$}} &
			{\parbox[c][1cm][c]{0.5cm}{\centering $B_1^{}$}} &
			{\parbox[c][1cm][c]{0.6cm}{\centering \begin{equation*} \frac {p_{\textnormal{N}, 1}^{}}{\varsigma_\textnormal{SNDR}^{}} \end{equation*} }} &
			{\parbox[c][1cm][c]{0.8cm}{\centering \begin{equation*} \frac {p_{\textnormal{IM3}, 1}^{}}{\varsigma_\textnormal{SNDR}^{}} \end{equation*} }} &
			{\parbox[c][1cm][c]{0.6cm}{\centering \begin{equation*} \frac {F_1^{}}{\varsigma_\textnormal{SNDR}^{}} \end{equation*} }} &
			{\parbox[c][1cm][c]{1.5cm}{\centering \begin{equation*} \sqrt { \varsigma_\textnormal{SNDR}^{}} V_{\textnormal{IIP3}, 1}^2  \end{equation*} }} &
			{\parbox[c][1cm][l]{1.8cm}{\centering \begin{equation} \label{eq:scaling_law_SNDR} \varsigma_{\textnormal{P}}^{} = \varphi \varsigma_\textnormal{SNDR}^{3/2} \end{equation}}} &
			$\begin{array} {rl} \varphi < 1, & 0 < \varsigma_{\textnormal{SNDR}}^{} < 1 \\ \varphi \geq 1, & 1 \leq \varsigma_{\textnormal{SNDR}}^{} < F_1 \end{array}$ \\
			$\mathit{SNDR}_1$ &
			{\parbox[c][1cm][c]{0.5cm}{\centering $\varsigma_{\textnormal{S}}^{} p_{\textnormal{S}, 1}^{}$}} &
			{\parbox[c][1cm][c]{0.5cm}{\centering $p_{\textnormal{I}, 1}^{}$}} &
			{\parbox[c][1cm][c]{0.5cm}{\centering $B_1^{}$}} &
			{\parbox[c][1cm][c]{0.6cm}{\centering $\varsigma_{\textnormal{S}}^{} p_{\textnormal{N}, 1}^{} $}} &
			{\parbox[c][1cm][c]{0.8cm}{\centering $\varsigma_{\textnormal{S}}^{} p_{\textnormal{IM3}, 1}^{} $ }} &
			{\parbox[c][1cm][c]{0.6cm}{\centering $ \varsigma_{\textnormal{S}}^{} F_1^{}$}} &
			{\parbox[c][1cm][c]{1.5cm}{\centering \begin{equation*}  \frac{V_{\textnormal{IIP3}, 1}^2}{\sqrt {\varsigma_\textnormal{S}^{}}}   \end{equation*} }} &
			{\parbox[c][1cm][l]{1.8cm}{\centering \begin{equation} \label{eq:scaling_law_ps} \varsigma_{\textnormal{P}}^{} = \varphi \varsigma_\textnormal{S}^{-3/2} \end{equation}}} &
			$\begin{array} {rl} \varphi < 1, & \varsigma_{\textnormal{S}}^{} > 1 \\ \varphi \geq 1, & \frac{1}{F_1} < \varsigma_{\textnormal{S}}^{} \leq 1 \end{array}$ \\
			$\mathit{SNDR}_1$ &
			{\parbox[c][1cm][c]{0.5cm}{\centering $p_{\textnormal{S}, 1}^{}$}} &
			{\parbox[c][1cm][c]{0.5cm}{\centering $\varsigma_{\textnormal{I}}^{} p_{\textnormal{I}, 1}^{}$}} &
			{\parbox[c][1cm][c]{0.5cm}{\centering $B_1^{}$}} &
			{\parbox[c][1cm][c]{0.6cm}{\centering $p_{\textnormal{N}, 1}^{}$}} &
			{\parbox[c][1cm][c]{0.8cm}{\centering $p_{\textnormal{IM3}, 1}^{}$}} &
			{\parbox[c][1cm][c]{0.6cm}{\centering $F_1^{}$}} &
			{\parbox[c][1cm][c]{1.5cm}{\centering $\varsigma_{\textnormal{I}}^{3/2} V_{\textnormal{IIP3}, 1}^2$}} &
			{\parbox[c][1cm][l]{1.8cm}{\centering \begin{equation} \label{eq:scaling_law_interference} \varsigma_{\textnormal{P}}^{} = \varsigma_\textnormal{I}^{3/2} \end{equation} }} &
			$\varphi = 1$ \\ 
			\noalign{\hrule height 2pt}
			
		\end{tabular}			
	\end{table*}
	Expression \eqref{eq:original_power_ratio} is a universal tool for calculating front end power scaling and can be used for all application-environment scenarios, under the condition that the corresponding front ends are implementable.
	However, one does need to use \eqref{eq:original_power_ratio} in a careful and structured way due to interdependencies between the fundamental parameters $(\mathit{SNDR}, B, p_\textnormal{S}^{}, p_\textnormal{I}^{})$ and noise-distortion ratio $\alpha_{\textnormal{IM3}}^{}$, system-level design parameters $(p_\textnormal{N}^{}, p_\textnormal{IM3}^{}, p_\textnormal{IIP3}^{})$ and circuit-level parameters $(F_\textnormal{AFE}^{}, V_{\textnormal{IIP}3}^2)$.
	More specifically, for a particular scenario, $\mathit{SNDR}$, $B$, $p_\textnormal{S}^{}$, $p_\textnormal{I}^{}$ and $\alpha_{\textnormal{IM3}}^{}$ will through \eqref{eq:IIP_pi_connection}, \eqref{eq:alpha_def} and \eqref{eq:SNDR_definition} yield $p_\textnormal{N}^{}$, $p_\textnormal{IM3}^{}$ and $p_\textnormal{IIP3}^{}$, which through \eqref{eq:noise_power_F_connection} and \eqref{eq:IIP3_voltage_power_connection} result in $F_\textnormal{AFE}^{}$ and $V_{\textnormal{IIP}3}^2$.
	Combining $F_\textnormal{AFE}^{}$ from pre- and post-scaling scenarios yields $\varphi$ from \eqref{eq:phi_definition}, noise-distortion ratios $\alpha_{\textnormal{IM3}}^{}$ give the value of $\delta$ from \eqref{eq:delta_def}, and the values of fundamental parameters result in respective scaling ratios, all of which is combined in \eqref{eq:original_power_ratio} for the final result.
	
	In order to isolate the scaling of power with only one of the fundamental parameters, we assume that the value of the parameter in question scales between the scenarios while other parameters remain constant.
	In this way, we obtain a restricted set of application-environment scenarios with high practical relevance, examined in detail in Section \ref{sec:ramifications}.
	Additionally, in all scenarios it is assumed that pre- and post-scaling $\alpha_{\textnormal{IM3}}^{}$ values are the same, i.e. that input-referred thermal noise and third-order distortion levels are kept at a constant ratio.
	Available literature on systematic receiver design suggests that in practice, the value of $\alpha_{\textnormal{IM3}}^{}$ is chosen to be small (typically on the order of 0.1) so that the third-order distortion is much weaker than the thermal noise, with the choice being consistent over different application-performance scenarios \cite[Ch. 13]{Razavi2011}.
	This consistency over scenarios is in line with our constant-$\alpha_{\textnormal{IM3}}^{}$ assumption.		
		
	The laws describing the scaling of front end power with fundamental parameters are given in Table~\ref{tab:scaling_laws}, expr. \eqref{eq:scaling_law_B}-\eqref{eq:scaling_law_interference}.
	Each of the four rows of the table, corresponding to a particular scaling law, also provides a comprehensive list of application-environment constraints (columns 1-4), together with a list of resulting system/circuit design requirements (columns 4-8) needed for scaling laws to hold in practical implementations, obtained through \eqref{eq:noise_power_F_connection}-\eqref{eq:IIP_pi_connection}, as discussed above.
	Note that the constraints on the scaling of $B$ also double as explicit design requirements.
						
	The four scaling laws can be stated in dB domain in form of convenient rules of thumb, as follows:
	\begin{enumerate}
		\item (\emph{Power consumption-bandwidth scaling law}): For every 1 dB increase/decrease of system bandwidth, the power consumption of an optimally designed analog front end increases/decreases by at least 1 dB.
		\begin{itemize}
			\item It is well known that the power consumption of standard analog blocks scales linearly with bandwidth \cite{Svensson2015}.
			This scaling law demonstrates that the linear power-bandwidth relation extends also to a chain of analog blocks.
		\end{itemize}
		
		\item (\emph{Power consumption-$\mathit{SNDR}$ scaling law}): For every 1 dB increase/decrease of $\mathit{SNDR}$, the power consumption of an optimally designed analog front end increases/decreases by at least 1.5 dB.
		\begin{itemize}
			\item This novel scaling law serves as a fundamental relation for analyzing power-performance tradeoffs in analog receiver design, as analyzed more in-depth in Sections \ref{sec:uncoded} and \ref{sec:coding}.
		\end{itemize}
		
		\item (\emph{Power consumption-received power scaling law}): For every 1 dB increase/decrease of received wanted signal power, the power consumption of an optimally designed analog front end decreases/ increases by at least 1.5 dB.
		\begin{itemize}
			\item This relation will be useful in analyzing power savings of a front end that adapts to a fluctuating received signal level while maintaining constant performance, as will be presented in Section \ref{sec:fading_adaptation}.
		\end{itemize}
		
		\item (\emph{Power consumption-interference scaling law}): For every 1 dB increase/decrease of the out-of-band interference power, the power consumption of an optimally designed analog front end increases/decreases by 1.5 dB.
		\begin{itemize}
			\item By defining the signal-to-interference ratio $\mathit{SIR} =  {p_{\textnormal{S}}^{}}/{p_{\textnormal{I}}^{}}$, this scaling can be reformulated as
			$\varsigma_{\textnormal{P}}^{} = \varsigma_{\textnormal{SIR}}^{-3/2}$,	where $\varsigma_{\textnormal{SIR}}^{}$ is the scaling of the $\mathit{SIR}$.
			An identical scaling law was presented in \cite{Heuvel2014}, where $P_{\textnormal{AFE}}^{}$ was optimized for energy efficiency. 	
			Scaling law \eqref{eq:scaling_law_interference} is of importance when analyzing the power consumption of a front end that dynamically adapts its linearity to the interference level while maintaining constant performance.
			A detailed theoretical analysis of such a front end will be given in Section \ref{sec:interference_adaptation}.
		\end{itemize}
	\end{enumerate}
	
	Laws \eqref{eq:scaling_law_B} - \eqref{eq:scaling_law_ps} are characterized by the fact that the underlying scaling asks for tuning of the noise figure, which in turn makes the parameter $\varphi$ scaling-dependent.	
	More specifically, for $\varsigma_*^{}$, where $* \in (\textnormal{B}, \textnormal{SNDR})$,
	\begin{equation}
		\varphi = \frac {F_{\textnormal{AFE}, 1} - 1}{F_{\textnormal{AFE}, 1} - \varsigma_*^{}},
	\end{equation}
	whereas for $\varsigma_{\textnormal{S}}^{}$ we have
	\begin{equation}
	\varphi = \frac {F_{\textnormal{AFE}, 1} - 1}{F_{\textnormal{AFE}, 1} - \frac {1}{\varsigma_\textnormal{S}^{}}}.
	\end{equation}
	The dependence of $\varphi$ on the scaling parameters is outlined in the last column of Table \ref{tab:scaling_laws}.
	The constraint $\varphi > 0$, i.e. the fact that it is physically impossible to have a front end with $F < 1$ imposes theoretical limitations on the values of scaling $\varsigma$.
	Furthermore, dependence of $\varphi$ on $\varsigma$ causes deviations from the ideal scaling of power (linear with bandwidth or following the 3/2 power law in case of $\mathit{SNDR}$ and received power).
	In order to have proper scaling laws, it is necessary for $\varphi$ to be independent of $\varsigma$.
	This condition is approximately satisfied in two cases:
	\begin{itemize}
		\item $F_{\textnormal{AFE}, 1} \gg 1 \Rightarrow \varphi \approx 1$;
		\item $\varsigma_\textnormal{B}^{}, \varsigma_\textnormal{SNDR}^{} \ll 1~\text{or}~\varsigma_{\textnormal{S}}^{} \gg 1 \Rightarrow \varphi \approx \frac {F_{\textnormal{AFE}, 1} - 1}{F_{\textnormal{AFE}, 1}}$.
	\end{itemize}
	
	At first, it can appear that the set of scenarios in which power scaling laws \eqref{eq:scaling_law_B}-\eqref{eq:scaling_law_ps} are close to ideal ($\varphi \approx 1$) is based on such a restrictive sequence of assumptions that their practical relevance is questionable.	
	However, a closer look reveals that all the assumptions we used are commonplace in practice and/or of high practical interest for low-power design.
	To start with, $F_{\textnormal{AFE}, 1} \gg 1$ is typical for worst-case front end designs with a large OOB blocking signal present \cite[Ch. 13]{Razavi2011}, \cite{Borremans2011}.
	Moreover, radical scaling down of system bandwidth (e.g. going from a wideband to a narrowband system), drastic downscaling of $\mathit{SNDR}$ requirement (due to e.g. use of power-efficient transmission techniques) or adaptation to wanted signal power that becomes much larger than worst-case (reference sensitivity) due to fading fluctuations are all use-cases of interest for low-power applications \cite{Xu2018}, \cite{Sen2016}.
	

	\section{Ramifications of the scaling laws}
	\label{sec:ramifications}
	
	The scaling laws presented in the previous section constitute a set of tools which prove to be very useful in the design of receivers where power consumption is of high importance.	
	Namely, as the laws in the preceding section formally show, the power consumption of the analog front end can be lowered by using one (or more) of the following techniques:
	\begin{itemize}
		\item Intentionally degrading the bit/symbol error rate ($\mathit{SER}$), which consequently reduces the $\mathit{SNDR}$ requirement;
		\item Keeping $\mathit{BER}$ or $\mathit{SER}$ constant while applying some transmission technique that allows for lower $\mathit{SNDR}$ (e.g. use of error control coding);	
		\item Keeping the $\mathit{SNDR}$ constant while making the AFE reconfigurable so that it adapts to the changes in the environment (e.g. fading level fluctuations, OOB interference level).
	\end{itemize}
	
	The scaling laws serve as a basis for estimates of the extent of power savings that can be achieved in the AFE if the aforementioned techniques are applied.
	System designers can then decide on which techniques to incorporate in their systems, and hardware designers are provided with general guidelines on how to increase the power efficiency of circuit designs.
	
	\subsection{Preliminaries: limitations on hardware relaxation}
	\label{subsec:preliminaries}
	
	Throughout the analysis that follows, we consider analog front ends designed for different target values of noise and distortion.			
	When it comes to realistic hardware designs, however, it is reasonable to assume that the range of these values is limited.
	Naturally, there are fundamental physical constraints on the minimum noise (or distortion) level that a circuit can deliver, but, equally important, there are also upper bounds, imposed by either functionality or technology process constraints \cite{Svensson2015}.	
	Hence, in line with considerations from the previous section, we establish a permissible tuning range $\mu$ that applies to both noise figure and IP3.
	It is defined as the value of scaling of noise/linearity for which, given all architectural and physical limitations, the following holds:
	\begin{itemize}
		\item The noise figure $F_\textnormal{AFE}$ can be degraded from the reference value $F_{\textnormal{AFE}, 1}$ to a maximum value of $F_{\textnormal{AFE}, 2} = \mu F_{\textnormal{AFE}, 1}$,
		\item It is possible to degrade IP3 from the reference value $V_{\textnormal{IIP3}, 1}^2$ to a minimum possible value of $V_{\textnormal{IIP3}, 2}^2 = \frac {1}{\sqrt{\mu}} V_{\textnormal{IIP3}, 1}^2$.
	\end{itemize}
	

	\subsection{Power- and energy-efficient AFEs through intentional degradation of performance, uncoded case}
	\label{sec:uncoded}
	
	In this section, we focus on systems using M-QAM without any error control coding.
	With the aim of saving power, System 2 either uses a lower QAM constellation order $M$ or operates at a higher symbol error probability $P_{\textnormal{e}}$, formally, $M_2 \leq M_1$ or $P_{\textnormal{e}, 2} \geq P_{\textnormal{e}, 1}$.
	As indicated in Section \ref{sec:scaling_laws}, the two systems are otherwise assumed to use the same bandwidth (and thus the same symbol rate $R_{\textnormal{s}}$), are affected by same OOB interference level and experience the same wanted signal power.
	
	We assume that the classical matched-filter detector is employed at the receiver.
	If the thermal Gaussian noise dominates the IM3, i.e. $\alpha_{\textnormal{IM3}}^{} \ll 1$, the matched-filter receiver is optimal in the sense of maximum aposteriori detection.
	Under these circumstances, an upper bound on $\mathit{SER}$ for a square M-QAM ($M = 2^{2k},~k \in \mathbb{N}$) can be determined \cite{Proakis2014}, which yields the inequality
	\begin{equation}
	\label{eq:critical_SNDR}
	\mathit{SNDR} \leq \rho \frac {M-1}{3 \log_2 M} \left [  Q^{-1} \left (  \frac{P_\textnormal{e}}{4}  \right )  \right ]^2, 
	\end{equation}
	where $\rho = R_{\textnormal{b}}/B$ is the spectral efficiency of the uncoded system ($R_{\textnormal{b}}$ is the information bitrate) and $Q^{-1}(\cdot) $ the inverse of the upper tail probability function of a unit-variance Gaussian random variable.
	At high $\mathit{SNDR}$s, the upper bound in \eqref{eq:critical_SNDR} is tight.
	
	We proceed by constructing a ratio of the upper bounds from \eqref{eq:critical_SNDR} that apply to the two distinct scenarios under analysis.
	This ratio is given as
	\begin{equation}
	\label{eq:achievable_gamma_uncoded_QAM}
	\varsigma_{\textnormal{SNDR}}^{} \geq \frac {M_2 - 1}{M_1 - 1} \left [ \frac {Q^{-1} \left ( P_{\textnormal{e}, 2}/4  \right )}{Q^{-1} \left ( P_{\textnormal{e}, 1}/4  \right )} \right ]^2,
	\end{equation}
	where the fact that $B$ is the same for the two systems is used.
	Taking into account the practical limits on noise/linearity scaling, discussed in Section \ref{subsec:preliminaries}, along with law \eqref{eq:scaling_law_SNDR}, the achievable scaling of front end power, $\varsigma_{\textnormal{P, a}}^{}$, is found to be
	\begin{equation}
	\label{eq:achievable_power_scaling}
		\varsigma_{\textnormal{P, a}}^{} < \max \{ \varsigma_{\textnormal{SNDR}}^{3/2}, \mu^{-3/2} \}.
	\end{equation}	
	By combining this together with \eqref{eq:achievable_gamma_uncoded_QAM} and the fact that the slack of the $\mathit{SER}$ upper bound increases with decreasing $\mathit{SNDR}$, we obtain the upper bound on the achievable AFE power downscaling:	
	\begin{equation}
	\label{eq:power_scaling_QAM}
	\varsigma_{\textnormal{P, a}}^{} \leq \max \left \{ \left ( \frac {M_2 - 1}{M_1 - 1}  \right )^{3/2} \left [ \frac {Q^{-1} \left ( P_{\textnormal{e}, 2}/4  \right )}{Q^{-1} \left (P_{\textnormal{e}, 1}/4  \right )} \right ]^3, \mu^{-3/2}  \right \}.
	\end{equation}
	In other words, the AFE power can be decreased by \textit{at least} the value of the right hand side of \eqref{eq:power_scaling_QAM}. 	
	For large $\mathit{SNDR}$s and large $F_{\textnormal{AFE}, 1}^{}$, the bound is tight.	
	
	\begin{figure}
		\centering      
		\includegraphics[width=0.6\textwidth]{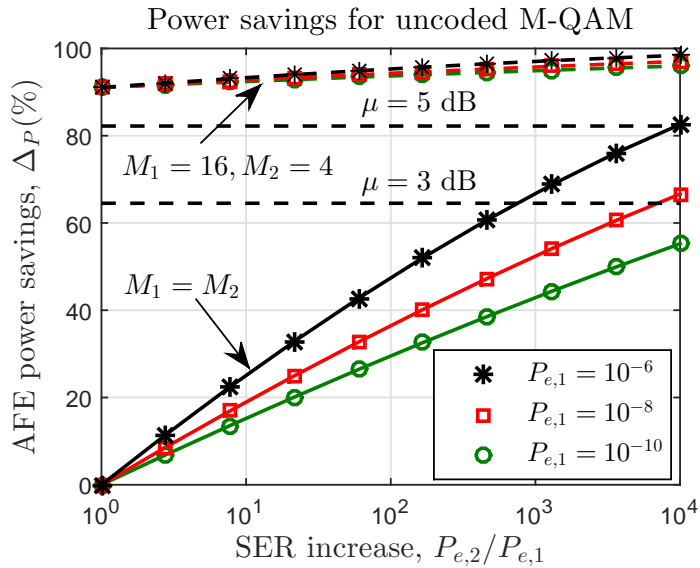}          
		\caption{Savings in AFE power consumption when symbol error probability and/or constellation order are degraded, for uncoded square QAM. For limited flexibility AFEs, the savings cap at values indicated by horizontal dashed lines.}
		\label{fig:uncoded_QAM}
	\end{figure}
	The obtained bound enables the derivation of laws describing the performance-power consumption tradeoff in systems using uncoded QAM when there are no limits on $\mathit{SNDR}$ tuning, $\mu \to \infty$.	
	In one case, we keep $P_\textnormal{e}$ constant but reduce the number of bits per symbol $b = \log_2 M$ by $\Delta_\textnormal{b}^{} = b_1 - b_2$.
	This yields
	\begin{equation}
	\label{eq:uncoded_QAM_bitrate_scaling}
	\varsigma_\textnormal{P}^{} < \left ( \frac {M_2}{M_1}  \right )^{3/2} = 2^{-\frac {3}{2} \Delta_\textnormal{b}^{}}.
	\end{equation}
	Therefore, the power consumption of an infinitely flexible AFE decreases at least exponentially with the difference in bits/symbol, or equivalently, with the difference in raw uncoded bitrate.	
	In another setting, we assume $M$ is the same between the two systems but target $\mathit{SER}$ is increased from $P_{\textnormal{e}, 1}^{}$ to $P_{\textnormal{e}, 2}^{}$.
	Using the bound $Q(x) \leq e^{-x^2/2}$, we get
	\begin{equation}
	\varsigma_\textnormal{P}^{} \leq \left [ \frac {Q^{-1} \left ( P_{\textnormal{e}, 2}^{}/4  \right )}{Q^{-1} \left ( P_{\textnormal{e}, 1}^{}/4  \right )} \right ]^3 
	\leq \left ( \frac {1 - 1.66 \log_{10} P_{\textnormal{e}, 2}^{}}{1 - 1.66 \log_{10} P_{\textnormal{e}, 1}^{}}  \right )^{3/2}.
	\end{equation}
	Assuming additionally that the order of magnitude $\omega_\textnormal{e}^{}~=~\log_{10} P_\textnormal{e}^{}$ of $\mathit{SER}$ is low enough, we get
	\begin{equation}
	\label{eq:uncoded_QAM_SER_scaling}
	\varsigma_\textnormal{P}^{} \leq \left ( \frac {\omega_{\textnormal{e}, 2}^{}}{\omega_{\textnormal{e}, 1}^{}}  \right )^{3/2}.
	\end{equation}
	In other words, we can say that the power consumption of the AFE with infinite flexibility scales at least as $\mathcal{O} \left ( \omega^{3/2} \right )$.

	For convenience of presenting numerical results, we define the percentage savings of AFE power
	\begin{equation}
	\label{eq:savings_function}
	\Delta_{\textnormal{P}}^{} \triangleq 100 (1 - \varsigma_\textnormal{P}^{}) \quad [\%].
	\end{equation}
	These savings, represented in Fig.~\ref{fig:uncoded_QAM} imply that, if presented with a choice of whether to sacrifice bitrate or error rate in order to save power in the receiver, we should in general opt for the former.
	Taking into account hardware design limitations, substantial savings are achievable even when it is possible to scale down the $\mathit{SNDR}$ by as little as e.g. 3 dB; naturally, in order to harvest the full potential of the savings, the AFE should be made as flexible as hardware constraints permit.
	
	In order to provide a completely fair comparison between the systems, degradation of the performance and reduction of power consumption should be considered jointly.
	A joint metric for performance and power consumption is needed for this task, and one is readily found in the form of energy efficiency
	\begin{equation}
	\label{eq:energy_efficiency}
	 \eta_{\textnormal{AFE}}^{} \triangleq \frac {R_{\textnormal{b}}^{}}{P_{\textnormal{AFE}}^{}} \quad [ \textnormal{bits} / \textnormal{J}].
	\end{equation}	
	In the case when constellation size $M$ changes but error rate $P_{\textnormal{e}}$ stays fixed and with unlimited flexibility, the ratio of the two efficiencies is
	\begin{equation}
	\label{eq:efficiency_ratio}
		\frac {\eta_{\textnormal{AFE}, 2}^{}}{\eta_{\textnormal{AFE}, 1}^{}} = \frac {1}{\varsigma_\textnormal{P}^{}} \frac {R_{\textnormal{b}, 2}^{}}{R_{\textnormal{b}, 1}^{}} \geq \left ( \frac {M_1 - 1}{M_2 - 1}  \right )^{3/2} \frac {\log_2 M_2}{\log_2 M_1}.
	\end{equation}
	
	From here we can conclude that, for a fixed $P_{\textnormal{e}}$, $\eta_{\textnormal{AFE}}^{}$ will always improve if the size of the square QAM constellation is reduced.	
	As a quick proof, we consider the fact that for square QAM, $M = 2^{2k},~k \in \mathbb{N}$ and so for any $k > 1$ we have $2^{2k} - 1 > 1$.
	This also means that for any $k_1 > k_2, ~ k_1, k_2 \in \mathbb{N}$ it will hold that
	\begin{equation}
	\label{eq:identity_for_k_s}
		\left ( \frac {2^{2k_1} - 1}{2^{2k_2} - 1}  \right )^{3/2} \frac {k_2}{k_1} > 1.
	\end{equation}
	But the left hand side of (\ref{eq:identity_for_k_s}) is equivalent to the right hand side of (\ref{eq:efficiency_ratio}), which means that
	\begin{equation}
	\label{eq:energy_efficiency_uncoded_QAM_proof}
		\frac {\eta_{\textnormal{AFE}, 2}^{}}{\eta_{\textnormal{AFE}, 1}^{}} \geq 1
	\end{equation}
	for $M_2 < M_1$.
	
	Therefore, the smaller the QAM constellation, the more energy efficient the AFE of the receiver.
	We note that, in the case when $\eta$ is defined with respect to \emph{transmit signal power}, it is a well known fact that the energy efficiency increases with decreasing QAM constellation size \cite{Proakis2014}.
	With \eqref{eq:energy_efficiency_uncoded_QAM_proof}, however, we prove that this energy efficiency property of QAM constellations extends to the case of \emph{power consumption of analog receiver hardware}.
	
	\subsection{Power- and energy-efficient AFEs through use of error control coding}
	\label{sec:coding}
	\begin{table*}[t]
		\centering
		\small
		\caption{System parameters and theoretical power numbers for AFEs and decoders in systems using error control coding}
		\begin{tabular}{ c c c c c c c c }
			\hline
			\parbox[c][2.5cm][c]{1.2cm}{\centering Information\\bitrate [Mbps]} &
			 Code &
			 $r_\textnormal{c}$ &
			 \parbox[c][1.1cm][c]{2.6cm}{\centering $g_\textnormal{c}$ @ $\mathit{BER} = 10^{-3}$\\ {}[dB]} & $P_{\textnormal{AFE}}^{}$ [mW] &
			 $P_{\textnormal{dec}}^{}$ [mW] &
			 \parbox[c][1.1cm][c]{1.2cm}{\centering $P_{\textnormal{AFE}}^{} + P_{\textnormal{dec}}^{}$\\{} [mW]} &
			 \parbox[s]{1.3cm}{\centering Total\\energy efficiency [Gbits/J]} \\ \hline
			26.7 &
			\parbox[c][1cm][c]{1.8cm}{\centering uncoded} &
			- &
			- &
			35 (\textbf{ref. \cite{Abdulaziz2016}}) &
			0 &
			35 &
			0.76 \\
			13.35 &
			\parbox[c][1.3cm][c]{1.8cm}{\centering convolutional\\(7, 5)} &
			1/2 &
			3.1 (\textbf{ref. \cite{Studer2012}})&
			4.26 &
			0.56 (\textbf{ref. \cite{Studer2012}})&
			4.82 &
			2.77 \\
			8.89 &
			\parbox[c][1cm][c]{1.8cm}{\centering turbo\\N = 6144} &
			1/3 &
			6.1 (\textbf{ref. \cite{Belfanti2013}})&
			0.82 &
			8.3 (\textbf{ref. \cite{Belfanti2013}})&
			9.12 &
			0.96 \\ \hline
		\end{tabular}		
		\label{tab:coded_system_numbers}
	\end{table*}
	Error control coding (ECC) techniques are used to improve reliability (error rate performance) of communication systems when $\mathit{SNDR}$ is kept fixed.
	Seen from another angle, when the error rate is constrained to be the same for uncoded and coded systems, coding can be used to improve the power efficiency of communication systems as a consequence of relaxed requirements on $\mathit{SNDR}$.
	Here we analyze the case when this potential for increased power efficiency is used by the receiver (it can also be used by the transmitter, or be distributed between the two).
	
	Power efficiency gain of coded systems is usually expressed in terms of the coding gain $g_\textnormal{c}^{}$.
	By assuming that $\alpha_{\textnormal{IM3}}^{} \ll 1$, we can approximate the PSD of the sum of all impairments by additive white Gaussian noise PSD $N_0^{}$ and define the ratio $E_\textnormal{b}^{} / N_0^{}$ of energy per bit $E_\textnormal{b}^{}$ and $N_0$.
	Given the $E_\textnormal{b}^{} / N_0^{}$ values required to achieve the same error probability with and without coding, the coding gain is defined as
	\begin{equation}
		g_\textnormal{c}^{} \triangleq \frac { \left (  E_\textnormal{b}^{} / N_0^{}  \right )_{\textnormal{uncoded}}}{ \left (  E_\textnormal{b}^{} / N_0^{}  \right )_{\textnormal{coded}}}.
	\end{equation}
	
	For finding the achievable AFE power reduction, we need to connect the coding gain $g_c$ with the $\mathit{SNDR}$ downscaling $\varsigma_ {\textnormal{SNDR}}^{}$, where $\mathit{SNDR}_1$ corresponds to the uncoded system and $\mathit{SNDR}_2$ to the coded one.	
	We do this by assuming that the system bandwidth is equal for both systems, which is a reasonable assumption for all applications where bandwidth is a limited resource.
	Consequently, using ECC will reduce spectral efficiency from $\rho_{\textnormal{uncoded}}^{}$ to $\rho_{\textnormal{coded}}^{} = r_\textnormal{c}^{} \rho_{\textnormal{uncoded}}^{}$, where $r_\textnormal{c}^{}$ is the coding rate.
	We additionally use the fact that $E_\textnormal{b} / N_0 = \mathit{SNDR}/\rho$ to obtain
	\begin{equation}
	\label{eq:coding_gain_gamma_connection}
	\varsigma_{\textnormal{SNDR}}^{} = \frac {r_\textnormal{c}^{}}{g_\textnormal{c}}^{},
	\end{equation}	
	and the associated achievable AFE power reduction (cf. \eqref{eq:achievable_power_scaling}) is then given by
	\begin{equation}
	\label{eq:coding_AFE_relaxation}
		\varsigma_\textnormal{P, a}^{} < \max \left \{ \left ( \frac {r_\textnormal{c}^{}}{g_\textnormal{c}^{}}  \right )^{3/2}, \mu^{-3/2}  \right \}.
	\end{equation}
	
\begin{figure}
		\centering      
		\includegraphics[width=0.6\textwidth]{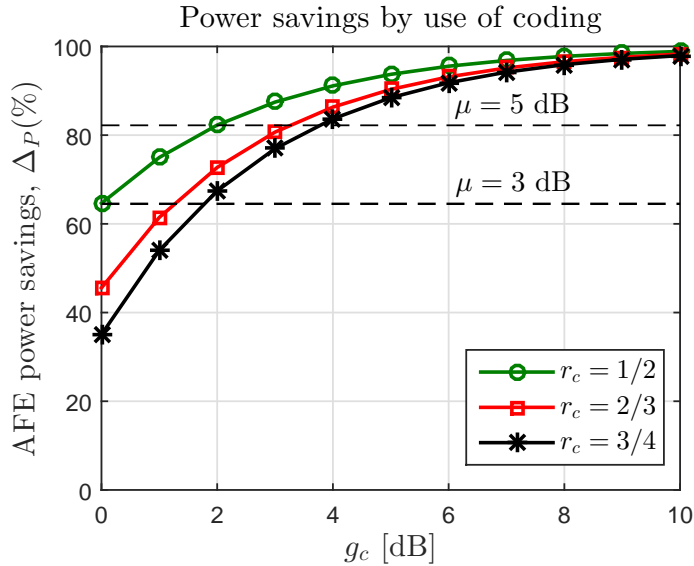}          
		\caption{Savings in AFE power consumption coming from use of error control coding. For limited flexibility AFEs, the savings cap at values indicated by horizontal dashed lines.}
		\label{fig:coded}
	\end{figure}

	The savings function (\ref{eq:savings_function}) for systems using coding is illustrated in Fig.~\ref{fig:coded}.
	An important observation to make here is that a large portion of the power savings (in absolute power terms) is harvested by low to intermediate coding gains.
	Additional absolute power savings that are brought about by employing stronger codes with larger coding gains are only marginal.
	This point is further elaborated in the follow-up.
	
	\subsubsection{Numerical example} here we provide a system design scenario that serves to illustrate the potential savings of AFE power consumption when ECC is used, and also to give some system-level design guidelines.		
	We assume a system with passband bandwidth of 40 MHz, BPSK modulation and single carrier transmission using raised cosine pulses with roloff of 0.5 over a flat-faded channel.
	Total receiver power (AFE + decoder) is calculated for three versions of the system: one uncoded and two with different types of ECC.
	If coding is used, the AFE design is relaxed accordingly.
	Power consumption values used here are ballpark quantities based on actual hardware designs.
	For the decoders, the power numbers obtained from the designs are modified to match the information bitrate (assuming that a linear extrapolation of decoder power consumption is possible at lower bitrates) and scaled to the same process (65 nm CMOS) and voltage (1.2 V).

	System parameters and calculated power numbers are listed out in Table \ref{tab:coded_system_numbers}.
	The use of coding allows for relaxation of the AFE by making it noisier and less linear, so its power consumption is ideally reduced as per (\ref{eq:coding_AFE_relaxation}).
	However, the overhead in power consumption stemming from the channel decoders also needs to be taken into account in order for the full story to be told.
	It can be seen that in the case of the system using convolutional codes (CC), a massive reduction of AFE power comes with a relatively small power overhead for the decoding.
	Using turbo codes allows for further reductions of AFE power compared to the CC case, but at a cost of a relatively high decoding power overhead, which is due to the iterative nature of the turbo decoder.
	Dividing the information bitrate with total power consumption yields the energy efficiency of the receiver, which indicates that coding indeed enables an improvement of the receiver energy efficiency, but the best strategy is to use ``light'' codes, with moderate coding gains and simpler decoders.	
	
	We note that the relation between error control coding and overall energy efficiency of the system is a long-standing research topic, examined both empirically and theoretically in, e.g., \cite{Howard2006} and \cite{Grover2011}.
	However, these papers analyze the combination of decoding power and \emph{transmit power}, whereas we focus on the \emph{total power of the receiver}, that is, the sum of decoding power and power consumed by supporting analog hardware.
	Here we have only touched upon this topic of high practical relevance, and a more thorough analysis is left for future work as it is out of the scope of this paper.
	
		\begin{table*}[t]
		\centering
		\small
		\caption{Noise tuning parameters and normalized power consumption of fading-adaptive front ends with limited adaptation range}
		\begin{tabular}{cccc}
			\hline
			\parbox[c][1cm][c]{3.1cm}{\centering Fading level } &
			$p_{\textnormal{N}}^{} (t)$ &
			$\frac {P_{\textnormal{AFE}}(t)^{}} {P_{\textnormal{AFE, wc}}^{}}$ &
			Remark \\ \hline
			\parbox[c][0.7cm][c]{3.1cm}{\centering $\phi (t) \leq \phi_{\textnormal{min}}^{}$} &
			$p_{\textnormal{N, min}}^{}$ &
			$=1$ &
			Outage \\
			\parbox[c][0.7cm][c]{3.1cm}{\centering $\phi_{\textnormal{min}} < \phi(t) \leq \mu \phi_{\textnormal{min}}^{} $ }&
			$\frac {\beta \phi(t)}{\left ( 1 + \alpha_{\textnormal{IM3}}^{}  \right ) \mathit{SNDR}_{\textnormal{min}}}$ &
			$< \left [ \frac {\phi (t)}{\phi_{\textnormal{min}}^{}} \right ]^{-3/2}$ &
			$\mathit{SNDR} = \mathit{SNDR}_{\textnormal{min}}^{}$ \\
			\parbox[c][0.7cm][c]{3.1cm}{\centering $\phi (t) > \mu \phi_{\textnormal{min}}^{} $} &
			$\mu p_{\textnormal{N, min}}^{}$ &
			$< \mu^{-3/2}$ &
			$\mathit{SNDR} > \mathit{SNDR}_{\textnormal{min}}^{}$ \\ \hline
		\end{tabular}		
		\label{tab:noise_tuning}
	\end{table*}
	As for the energy efficiency of the AFE alone, it can be quickly shown that it always improves with coding.
	This is done by setting up the ratio of energy efficiencies (\ref{eq:energy_efficiency}) for the coded and uncoded system, which gives
	\begin{equation}
	 \frac {\eta_{\textnormal{coded}}^{}}{\eta_{\textnormal{uncoded}}^{}} = \frac {1}{\varsigma_\textnormal{P}^{}} \frac {R_{\textnormal{b}, coded}^{}}{R_{\textnormal{b}, uncoded}^{}} = \frac {r_\textnormal{c}^{}}{\varsigma_\textnormal{P}^{}} > \frac {g_\textnormal{c}^{3/2}}{\sqrt{\varsigma_\textnormal{P}^{}}}
	\end{equation}
	in the case of infinite AFE flexibility.
	But the obtained ratio is always $> 1$ for $g_\textnormal{c}^{} > 1$ (for a properly designed code operating at a large enough $\mathit{SNDR}$).
		
	Overall, the results in this section lead to the conclusion that low power applications that harness error control coding gains for the goal of relaxing the receiver favor simple codes with modest coding gains and simpler decoders over more powerful codes that ask for more involved decoding algorithms.
	Another, more general design guideline is that the power budget for the channel decoder must fit into the margin opened up by relaxing the AFE if the goal is to reduce the overall receiver power consumption.
	If we, on the other hand, consider solely the AFE, it can be shown that coding always improves its energy efficiency.
	
	\subsection{Power-efficient AFEs through adaptation to fading}
	\label{sec:fading_adaptation}

	In this section, we assume a single carrier transmission over a frequency flat wireless channel.
	Due to fading, received power $p_\textnormal{S}^{}$ will be time varying and can be well described as a random process
	\begin{equation}
	p_{\textnormal{S}}^{} (t) = \beta \phi(t),
	\end{equation}
	where $\beta$ subsumes the transmit power, transmit and receive antenna gains, pathloss and large-scale fading, which are all assumed constant in this context.
	Additionally, $\phi(t) = |h(t)|^2$, where $h(t)$ is a zero-mean unit-variance complex Gaussian random process, i.e. the small-scale fading adheres to the common Rayleigh fading model.
	It is well known that $\phi(t)$ has an exponential pdf \cite{Goldsmith2005}
	\begin{equation}
	\label{eq:fading_power_pdf}
	f_\Phi (\phi) = e^{-\phi}, \quad \phi \geq 0.
	\end{equation}

	A common design parameter for wireless systems is the outage probability $\Omega$, defined as the probability that the normalized fading power $\phi$ falls below some minimum acceptable level $\phi_{\textnormal{min}}^{}$ \cite{Goldsmith2005},
	\begin{equation}
	\label{eq:outage_definition}
	\Omega \triangleq \int_0^{\phi_{\textnormal{min}}^{}} f_\Phi (\phi) d \phi.
	\end{equation}
	In conjunction with $\phi_{\textnormal{min}}^{}$, an outage $\mathit{SNDR}$ is usually defined, which represents the minimum $\mathit{SNDR}$ that provides acceptable performance.
	Using $\phi_{\textnormal{min}}^{}$ and $\mathit{SNDR}_\textnormal{min}^{}$, a minimum (worst-case) thermal noise level is calculated as
	\begin{equation}
	\label{eq:fixed_pn_def}
	 p_{\textnormal{N, min}}^{} = \frac {\beta \phi_{\textnormal{min}}^{}}{\left ( 1 + \alpha_{\textnormal{IM3}}^{}  \right ) \mathit{SNDR}_{\textnormal{min}}^{}}.
	\end{equation}
	Therefore, a minimum noise level $p_{\textnormal{N, min}}^{}$ and a minimum third-order distortion $p_\textnormal{IM3, min}^{}$ need to be delivered by the AFE at least at the time instants where $\phi(t) = \phi_{\textnormal{min}}^{}$.
	For all practical purposes, however, AFEs are built so that they deliver minimum noise and distortion \textit{all the time}.	
	Since the outage probability $\Omega$ is typically chosen to be quite low (for example, on the order of $10^{-2}$), this means that for the vast majority of time, $\mathit{SNDR}$ delivered by these worst-case designs will be much larger than $\mathit{SNDR}_{\textnormal{min}}^{}$ and performance far better than the minimum acceptable one.
	
	\begin{figure}[h!]
		\centering      
		\includegraphics[width=0.5\textwidth]{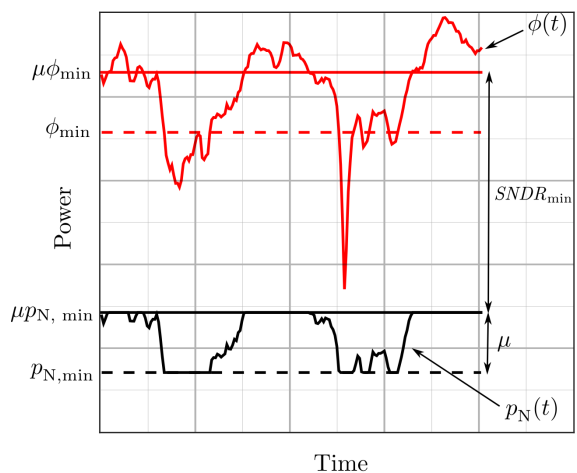}          
		\caption{Illustration of time-varying fading and system parameters for the fading-adaptive front end design}
		\label{fig:fading_illustration}
	\end{figure}
	Unless the variations in $\mathit{SNDR}$ are leveraged for increasing throughput (via adaptive modulation and coding), having the front end operate in a fixed manner represents a waste of power.
	If a fixed throughput and error rate are acceptable for a particular application, front end noise and linearity can be tuned to track the variations of received power and maintain constant $\mathit{SNDR}$ (effectively ``equalizing'' the channel).
	As indicated by results of Section \ref{sec:scaling_laws}, such an approach would result in a reduction of power consumed by the front end.
	
	We now turn to quantifying this reduction.
	Firstly, in line with considerations in Section \ref{subsec:preliminaries}, it is reasonable to assume that the noise level in an adaptive front end can be tuned only in a limited range $\left ( p_{\textnormal{N, min}}^{},~ \mu p_{\textnormal{N, min}}^{} \right )$ while being kept constant at the range boundaries for too small/large values of $\phi (t)$.
	The same logic extends to adapting the distortion level by means of tuning the nonlinearity, which yields the allowed range for the distortion of $\left ( p_{\textnormal{IM3, min}}^{},~ \mu p_{\textnormal{IM3, min}}^{} \right )$.
	The adaptation rule for thermal noise in a fading-adaptive front end with limited adaptation range is given in Table \ref{tab:noise_tuning}, with the most important parameters of interest illustrated in Fig.~\ref{fig:fading_illustration}.
	Using relations \eqref{eq:noise_power_F_connection} - \eqref{eq:IIP_pi_connection}, $\alpha_{\textnormal{IM3}, 1}^{} = \alpha_{\textnormal{IM3}, 2}^{}$ and the set of constraints from the third row of Table \ref{tab:scaling_laws}, these rules can be easily translated to feature circuit design parameters.
	
	
	\begin{figure*}[h!] 
		\centering
		\begin{tabular}{cc}
			\includegraphics[width=0.45\textwidth]{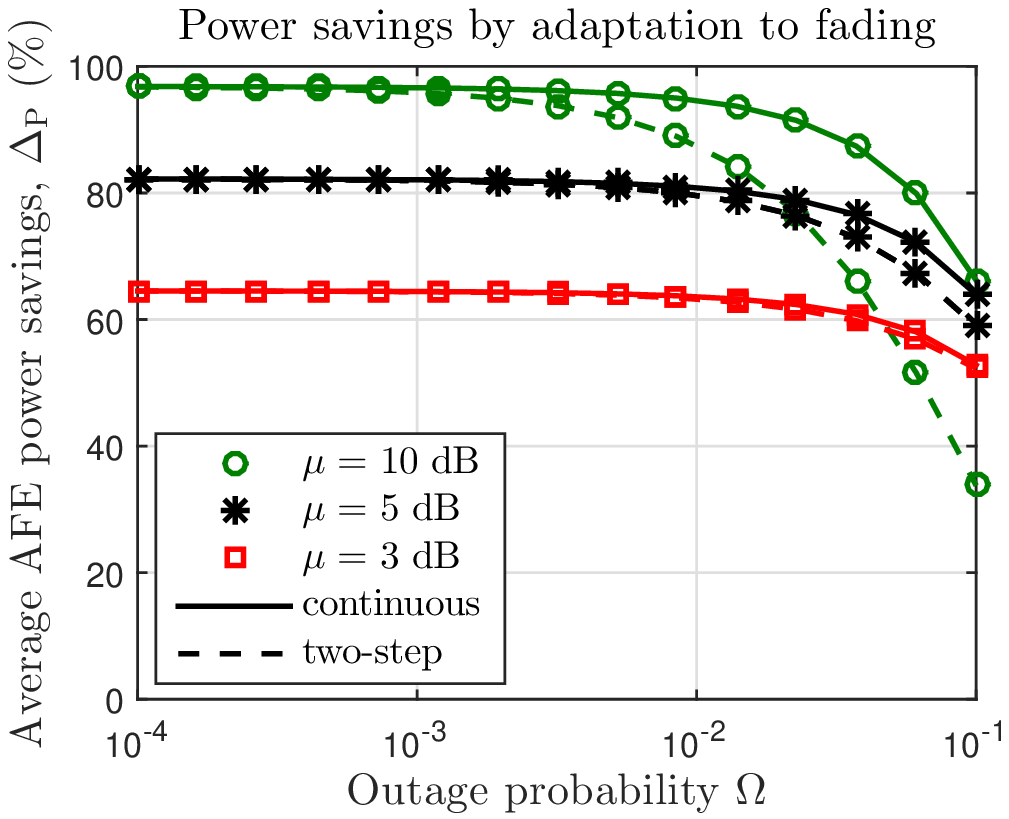} & \includegraphics[width=0.45\textwidth]{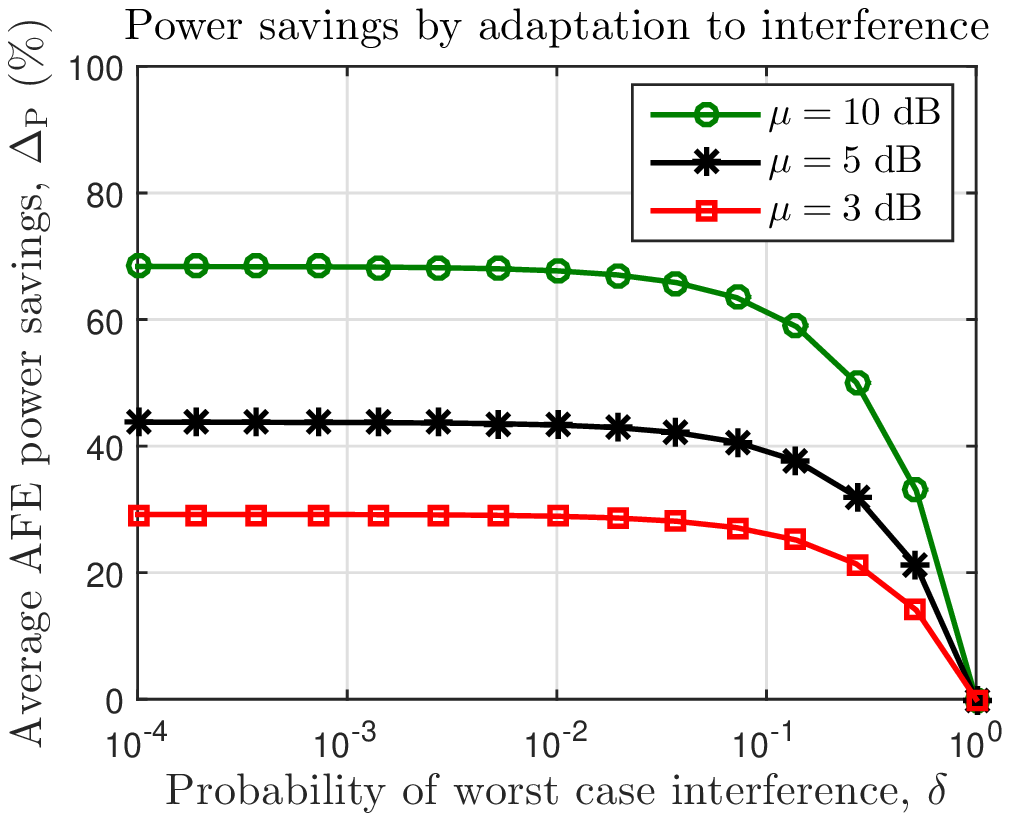} \\
			\parbox[c][1.5cm][t]{0.4\textwidth}{a) Average power savings of continuous and two-step fading-adaptive AFEs} &
			\parbox[c][1.5cm][t]{0.4\textwidth}{b) Average power savings of two-step interference-adaptive AFEs} \\
			\includegraphics[width=0.45\textwidth]{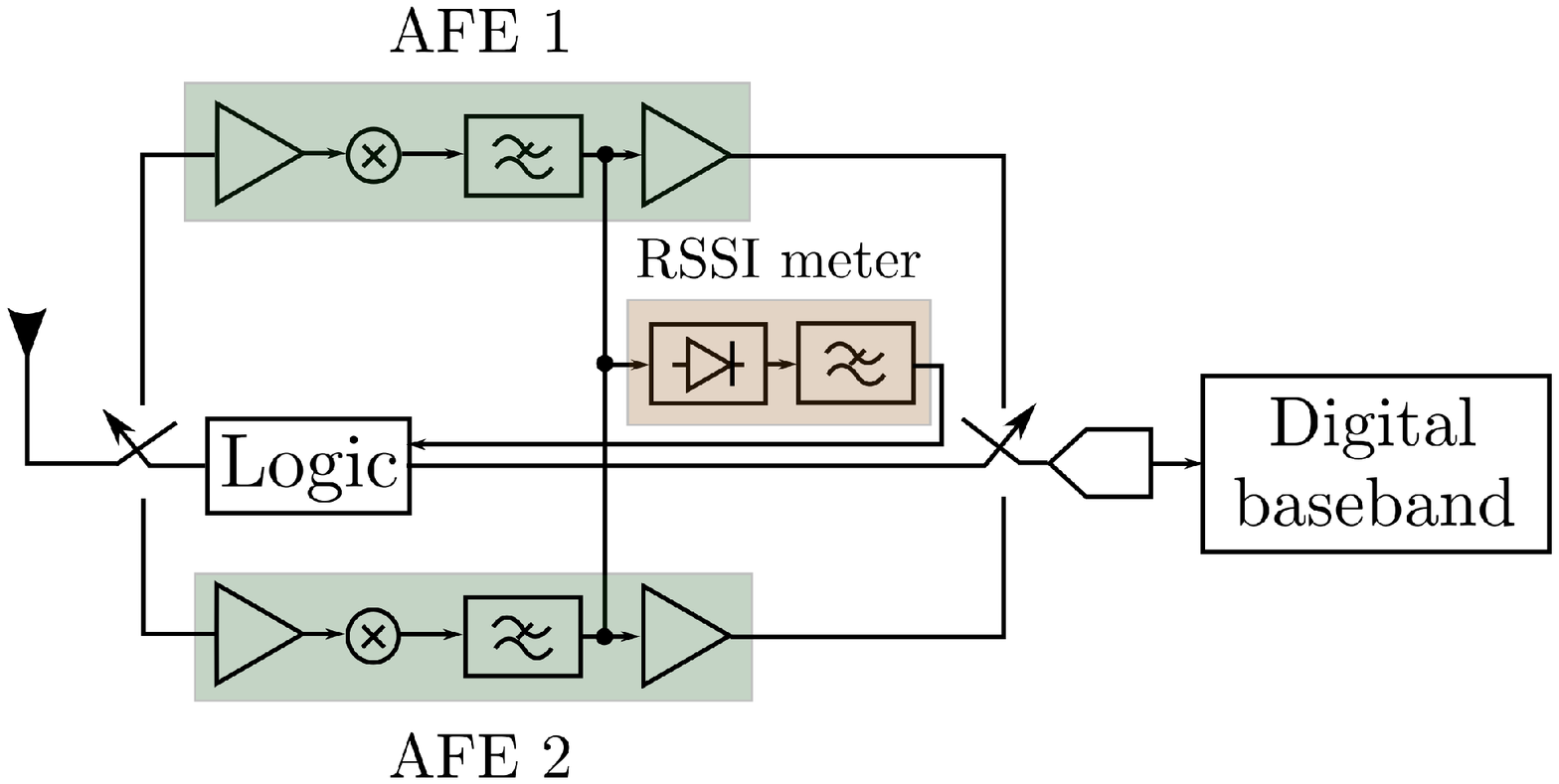} & \includegraphics[width=0.45\textwidth]{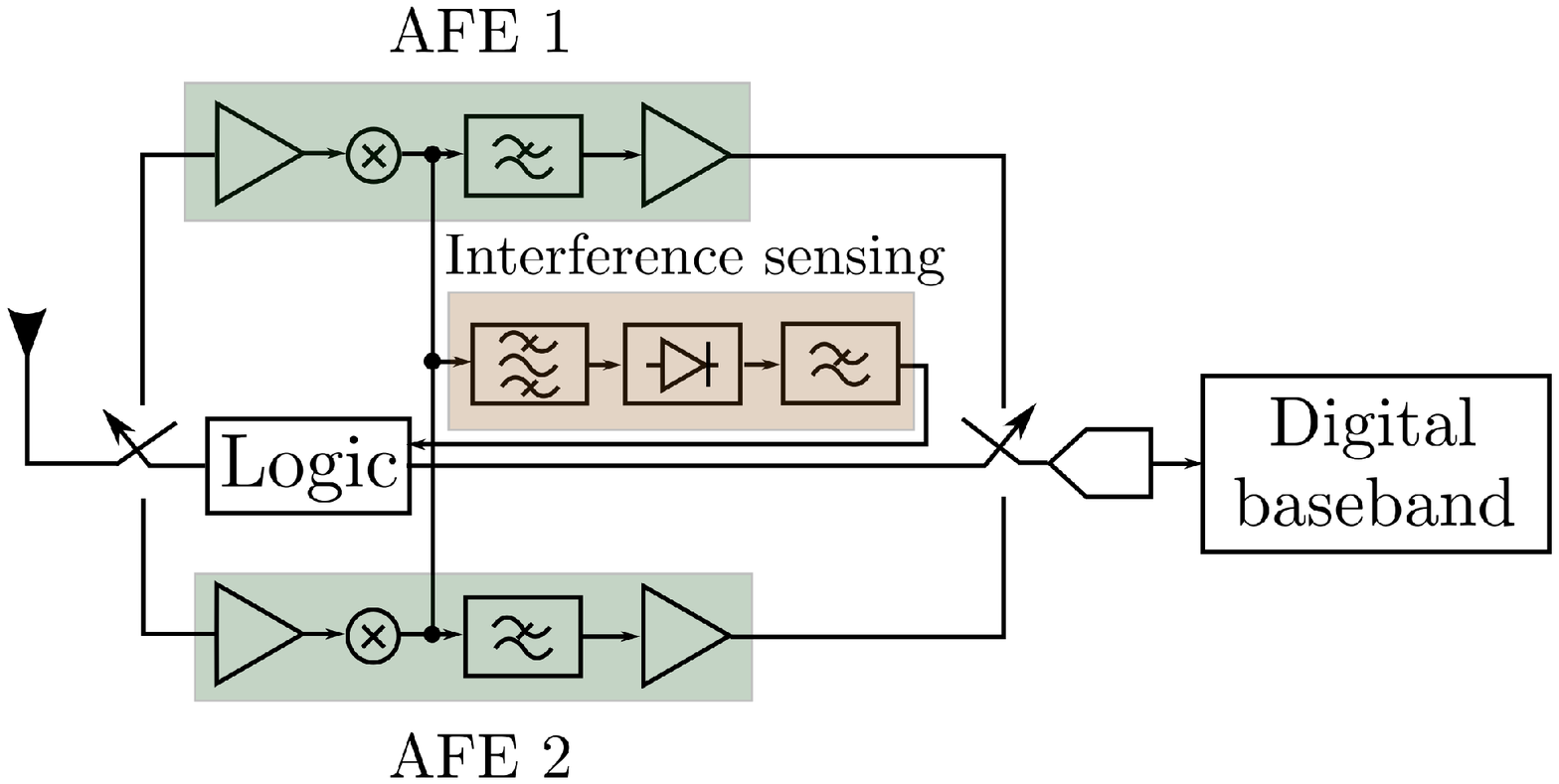} \\
			\parbox[c][2cm][t]{0.4\textwidth}{c) Example architecture for two-step adaptation to fading. Front end 1 has low noise figure and high IP3; front end 2 high noise figure and low IP3.} &
			\parbox[c][2cm][t]{0.4\textwidth}{d) Example architecture for two-step adaptation to interference. Front end 1 has high IP3; front end 2 low IP3.}
		\end{tabular}
		\caption{Theoretical power savings and conceptual illustrations of architectures for adaptive receivers}		
		\label{fig:adaptive_results_and_architectures}
	\end{figure*}

	We further denote by $P_{\textnormal{AFE, wc}}^{}$ the power consumption of the non-adaptive, worst-case front end architecture, designed to deliver $p_{\textnormal{N, min}}^{}$ and $p_{\textnormal{IM3, min}}^{}$ throughout.
	Taking into account scaling law \eqref{eq:scaling_law_ps}, the power consumption of the adaptive front end $P_{\textnormal{AFE}}^{}(t)$ normalized by $P_{\textnormal{AFE, wc}}^{}$ depends on $\phi (t)$ and is given in Table \ref{tab:noise_tuning}.
	From there, the expected value of power scaling $\varsigma_{\textnormal{P}}^{}$ for the adaptive front end can be calculated by assuming that $\phi (t)$ is an ergodic process (so time averages can be substituted by ensamble averages) as
	\begin{equation}
		\mathbb{E} \left \{  \varsigma_{\textnormal{P}}^{}  \right \} \leq \int_{0}^{\phi_{\textnormal{min}}^{}} e^{-\phi} \mathop{d \phi}~+~\phi_{\textnormal{min}}^{3/2} \int_{\phi_{\textnormal{min}}^{}}^{\mu \phi_{\textnormal{min}}^{}} \phi^{-3/2} e^{-\phi} \mathop{d \phi}
		+~\mu^{-3/2} \int_{\mu \phi_{\textnormal{min}}^{}}^{\infty} e^{-\phi} \mathop{d \phi},
	\end{equation}
	which yields 
	\begin{align}
		\label{eq:flexible_fading_adaptive_bound}
		\mathbb{E} \left \{  \varsigma_{\textnormal{P}}^{}  \right \}_{\textnormal{continuous}} &\leq 1 - e^{-\phi_{\textnormal{min}}} + \\ \nonumber
		&2 \left \{ \phi_{\textnormal{min}} e^{-\phi_{\textnormal{min}}} \left ( 1 - \frac{1}{\sqrt{\mu}} e^{(1 - \mu) \phi_{\textnormal{min}}  }  \right ) + \right. \\ \nonumber
		&\left. \phi_\textnormal{min}^{3/2} \left [ \Gamma \left ( \frac {1}{2}, \mu \phi_{\textnormal{min}}^{} \right ) - \Gamma \left ( \frac {1}{2},\phi_{\textnormal{min}}^{}  \right ) \right ] \right \} + \\ \nonumber
		&\mu^{-3/2} e^{-\mu \phi_{\textnormal{min}}},
	\end{align}	
	where $\Gamma (a, x)$ denotes the upper incomplete gamma function \cite{Abramowitz1965}.
	
	Achieving continuous tuning of noise and linearity can be challenging in practical implementations.
	Apart from adapting to the environment, the issue of random PVT (process, voltage, temperature) variations also needs to be accounted for.
	There exist solutions for jointly solving these practical problems, such as the one presented in 	\cite{Sen2012}, where LNAs with orthogonally tunable noise and linearity are combined with a simple online optimization algorithm, yielding substantial power savings.	
	An alternative way of tackling this issue is to form a bank of front ends that are optimally designed for different noise and linearity settings.
	During operation, the receiver would switch between different front ends based on the measured received power, keeping one front end active and switching off the rest.
	In the most basic case, such a bank would consist of only two front ends.
	A switching rule for this two-step adaptive front end that guarantees $\mathit{SNDR} \geq \mathit{SNDR}_\textnormal{min}$ can be defined as
	\begin{equation}
		p_\textnormal{N}^{} =
		\begin{cases}
			p_\textnormal{N, min}^{}, & \phi(t) \leq \mu \phi_{\textnormal{min}}^{} , \\
			\mu p_\textnormal{N, min}^{} , & \phi(t) > \mu \phi_{\textnormal{min}}^{} .
		\end{cases}
	\end{equation}
	Average power downscaling for the two-step front end is found to be
	\begin{equation}
	\label{eq:two_step_fading_adaptive_bound}
		\mathbb{E} \left \{  \varsigma_{\textnormal{P}}^{}  \right \}_{\textnormal{two-step}} \leq 1 - \left ( 1 - \mu^{-3/2}  \right ) e^{-\mu \phi_{\textnormal{min}}^{}}.
	\end{equation}

	Average power scaling for flexible and two-step front ends is converted to average savings as per \eqref{eq:savings_function} and shown in Fig.~\ref{fig:adaptive_results_and_architectures}a).
	When the tuning range $\mu$ is small, normalized signal power $\phi (t)$ is either in outage or above $\mu \phi (t)$ for most of the time, so continuous and two-step front ends have similar power savings.
	As the tuning range increases, more power can be saved, but in the case of large outage probability, $\phi(t)$ is rarely larger than $\mu \phi (t)$.
	This means that in the case of the two-step front end, the noisy, nonlinear, low power front end rarely gets activated and the power savings are significantly lower compared to continuous adaptation.
	In any case, the obtained savings are substantial \footnote{We reiterate that the front end power can be scaled down by \textit{at least} the values given by the right hand side of \eqref{eq:flexible_fading_adaptive_bound} and \eqref{eq:two_step_fading_adaptive_bound}, i.e. Fig.~\ref{fig:adaptive_results_and_architectures}a) illustrates a lower bound on possible savings!}, which should serve as a motivation for implementing fading-adaptive front ends in practice.
	In the case of two-step adaptation, such implementations can have an appealing simplicity.
	As means of illustration, we provide a high-level conceptual sketch of how they might look like, shown in Fig.~\ref{fig:adaptive_results_and_architectures}c).
	Under the condition that the channel select filter removes most of the OOB interference, the wanted signal power can be measured in the baseband by a simple power detector.
	This information, properly calibrated to account for in-band gains, can be used by a logic circuit which will drive the switching between the two front ends.
	
	\subsection{Power-efficient AFEs through adaptation to out-of-band interference}
	\label{sec:interference_adaptation}
	
	The analysis of practical implications of the AFE power scaling laws is concluded by looking into how much power can be saved if the AFE adapts its linearity to the OOB interferer level.
	It is assumed that the wanted signal, whose level does not change, is accompanied by two interferers with total power $p_\textnormal{I}^{}$ and equal, slowly time varying amplitudes, so that they can be well approximated by two tones.	
	
	We analyze a receiver structure that is able to adjust its linearity in two discrete steps and in doing so, adapt to the fluctuating interference level.	
	To this end, suppose that we have two analog front end designs at our disposal.
	One of them is designed for the worst-case interference level $p_{\textnormal{I, wc}}^{}$ (a value commonly prescribed in communication standards) and its linearity is equal to $p_{\textnormal{IIP}3, \textnormal{wc}}^{}$.
	On the other hand, the IP3 of the other design has been degraded down to the limits of implementability and is equal to $ p_{\textnormal{IIP}3, \textnormal{wc}}^{}/\sqrt{\mu}$ \footnote{
	This value is chosen in line with considerations from Section \ref{subsec:preliminaries} and provides a fair comparison with other results in this section.}.
	Otherwise, the bandwidth and noise figure of the two front ends are the same.
	
	The task of the receiver is to track the interference power and switch between the two front ends so that a minimum performance requirement is always satisfied, $\mathit{SNDR} \geq \mathit{SNDR}_{\textnormal{min}}^{}$, or equivalently, that the intermodulation distortion is always kept below a certain level:
	\begin{equation}
	\label{eq:IM3_constraint}
		p_{\textnormal{IM}3}^{} \leq p_{\textnormal{IM}3, \textnormal{wc}}^{} = \frac {p_{\textnormal{I, wc}}^{3}}{p_{\textnormal{IIP}3, \textnormal{wc}}^{2}}.
	\end{equation}
	Condition \eqref{eq:IM3_constraint} is met by a receiver which will tune its IP3 by switching between the described front ends in line with the following rule:
	\begin{equation}
		p_{\textnormal{IIP}3}^{} =
		\begin{cases}
			p_{\textnormal{IIP}3, \textnormal{wc}}^{}, & \frac{1}{\sqrt[3]{\mu}} p_{\textnormal{I, wc}}^{} < p_{\textnormal{I}}^{} \leq p_{\textnormal{I, wc}}^{}, \\
			\frac {1}{\sqrt{\mu}} p_{\textnormal{IIP}3, \textnormal{wc}}^{}, & p_{\textnormal{I}}^{} \leq \frac{1}{\sqrt[3]{\mu}} p_{\textnormal{I, wc}}^{},
		\end{cases}
	\end{equation}
	with one front end with desired linearity being on and the other one switched off.
	
	In order to characterize average power savings, it is not necessary to have the knowledge of the actual distribution of $p_\textnormal{I}^{}$.
	It is sufficient to assume that the probability of $p_\textnormal{I}^{} > p_\textnormal{I, wc}^{}$ is negligible (which is why this case is not covered by the adaptation rule), and that only the probability $\delta$ of interference being ``high'' is known, i.e.
	\begin{align}
	\label{eq:probability_high_interference}
		 \Pr \left \{  \frac {1}{\sqrt[3]{\mu}} p_{\textnormal{I, wc}}^{} < p_\textnormal{I}^{} \leq  p_{\textnormal{I, wc}}^{} \right \} &= \delta, \\ \nonumber	
		 \Pr \left \{  p_\textnormal{I}^{} \leq \frac {1}{\sqrt[3]{\mu}} p_{\textnormal{I, wc}}^{} \right \} &= 1 - \delta.
	\end{align}	
	As in the preceding section, we normalize the power consumption of the adaptive receiver with the power consumed by a non-adaptive receiver that utilizes only the high linearity front end. 
	By using \eqref{eq:scaling_law_interference}, we obtain
	\begin{equation}
		\frac {P_{\textnormal{AFE, adaptive}}}{P_{\textnormal{AFE, fix}}} =
		\begin{cases}
			1, &\frac{1}{\sqrt[3]{\mu}} p_{\textnormal{I, wc}}^{} < p_{\textnormal{I}}^{} \leq p_{\textnormal{I, wc}}^{}, \\
			\frac {1}{\sqrt{\mu}}, &p_{\textnormal{I}}^{} \leq \frac{1}{\sqrt[3]{\mu}} p_{\textnormal{I, wc}}^{},
		\end{cases}
	\end{equation}
	which, combined with \eqref{eq:probability_high_interference}, yields
	\begin{equation}
		\mathbb{E} \left \{  \varsigma_{\textnormal{P}}^{} \right \} = \delta + \frac {1 - \delta}{\sqrt{\mu}}.
	\end{equation}
	Average power savings of such a receiver are shown in Fig.~\ref{fig:adaptive_results_and_architectures}b).
	For example, given that $\mu = 10$ dB, the range of OOB interferer values for which the high linearity AFE is activated (worst-case interference) is $(0.46~p_{\textnormal{I, wc}}^{},~p_{\textnormal{I, wc}}^{})$.
	If the interference power is inside this range for 10\% of the time, the low linearity AFE would be used for the remaining 90\% of the time and the average power savings compared to a non-adaptive design are 60\%.
	Taking the ballpark power numbers for a front end from \cite{Abdulaziz2016}, this signifies a reduction of average front end power from 35 mW to 14 mW.
	Paper \cite{Abdulaziz2016} also suggests a practical implementation of the interference sensing circuit, consisting of a passband filter and an energy detector.
	We include this sensor in the high-level conceptual illustration of an interference-adaptive receiver, shown in Fig.~\ref{fig:adaptive_results_and_architectures}d).
	The sensor from \cite{Abdulaziz2016} consumes 10 mW, which combined with the reduced average AFE power consumption (and neglecting the consumption of the logic circuitry) yields 24 mW, which is still 30\% less than the power consumed by the non-adaptive receiver.
	
	
	\section{Conclusion}
	Based on a known result from circuit theory that has also been verified in practice, we determine scaling laws between performance and power consumption of an analog front end (AFE).
	The power consumption of the AFE is found to scale as $\mathit{SIR}^{-3/2}$ and at least as $\mathit{SNDR}^{3/2}$.
	These simple scaling laws can be used in a wide variety of communication-theoretic contexts, and some of the most important ones are explored.
	Namely, the power-SNR scaling law is extended to find the scaling laws between AFE power consumption and QAM constellation size, symbol error probability for QAM and error control coding gain and rate.
	Some general rules for low-power system design can be drawn from these laws: one example rule is that low-power applications favor ``light'' channel codes with moderate coding gains (such as simple convolutional codes) over more powerful ones, like turbo codes.
	Moreover, we derive laws that describe how front end power scales with environment parameters when performance is kept constant.
	Combined with fading and out-of-band blocker statistics, this enables us to determine theoretical average power savings of AFEs that adapt to the environment.
	The impressive results (about one order of magnitude reduction of power consumption in some cases) indicate that designing the front end so that it adapts to the environment is definitely a worthwhile effort.
	
	
	\section*{Acknowledgement}
	The authors would like to thank the Swedish Foundation for Strategic Research (SSF), which provided the funding of this research in the scope of the Digitally Assisted Radio Evolution (DARE) project.
	

\begin{thebibliography}{10}
		\providecommand{\url}[1]{#1}
		\csname url@samestyle\endcsname
		\providecommand{\newblock}{\relax}
		\providecommand{\bibinfo}[2]{#2}
		\providecommand{\BIBentrySTDinterwordspacing}{\spaceskip=0pt\relax}
		\providecommand{\BIBentryALTinterwordstretchfactor}{4}
		\providecommand{\BIBentryALTinterwordspacing}{\spaceskip=\fontdimen2\font plus
			\BIBentryALTinterwordstretchfactor\fontdimen3\font minus
			\fontdimen4\font\relax}
		\providecommand{\BIBforeignlanguage}[2]{{%
				\expandafter\ifx\csname l@#1\endcsname\relax
				\typeout{** WARNING: IEEEtran.bst: No hyphenation pattern has been}%
				\typeout{** loaded for the language `#1'. Using the pattern for}%
				\typeout{** the default language instead.}%
				\else
				\language=\csname l@#1\endcsname
				\fi
				#2}}
		\providecommand{\BIBdecl}{\relax}
		\BIBdecl
		
		\bibitem{Abidi2000}
		A. A. Abidi, G. J. Pottie and W. J. Kaiser, ``Power-conscious design of wireless circuits and systems,'' in Proceedings of the IEEE, vol. 88, no. 10, pp. 1528-1545, Oct. 2000.
		
		\bibitem{Svensson2015}
		C. Svensson, ``Towards power centric analog design,'' in IEEE Circuits and Systems Magazine, vol. 15, no. 3, pp. 44-51, 2015.
		
		\bibitem{Sheng2006}
		W. Sheng, A. Emira and E. Sanchez-Sinencio, ``CMOS RF receiver system design: a systematic approach,'' in IEEE Transactions on Circuits and Systems I: Regular Papers, vol. 53, no. 5, pp. 1023-1034, May 2006.
		
		\bibitem{Baltus2000}
		P. G. M. Baltus and R. Dekker, ``Optimizing RF front ends for low power,'' in Proceedings of the IEEE, vol. 88, no. 10, pp. 1546-1559, Oct. 2000.
		
		\bibitem{Heuvel2014}
		J. H. C. van den Heuvel, Y. Wu, P. G. M. Baltus, J. P. P. M. G. Linnartz and A. H. M. van Roermund, ``Front end power dissipation minimization and optimal transmission rate for wireless receivers,'' in IEEE Transactions on Circuits and Systems I: Regular Papers, vol. 61, no. 5, pp. 1566-1577, May 2014.
		
		\bibitem{Meghdadi2014}
		M. Meghdadi and M. Sharif Bakhtiar, ``Two-dimensional multi-parameter adaptation of noise, linearity, and power consumption in wireless receivers,'' in IEEE Transactions on Circuits and Systems I: Regular Papers, vol. 61, no. 8, pp. 2433-2443, Aug. 2014.
		
		\bibitem{Do2010}
		A. V. Do, C. C. Boon, M. A. Do, K. S. Yeo and A. Cabuk, ``An energy-aware CMOS receiver front end for low-power 2.4-GHz applications,'' in IEEE Transactions on Circuits and Systems I: Regular Papers, vol. 57, no. 10, pp. 2675-2684, Oct. 2010.
		
		\bibitem{Hueber 2008}
		G. Hueber, J. Zipper, R. Stuhlberger and A. Holm, ``An adaptive multi-mode RF front-end for cellular terminals,'' 2008 IEEE Radio Frequency Integrated Circuits Symposium, Atlanta, GA, 2008, pp. 25-28.		
		
		\bibitem{Senguttuvan2008}
		R. Senguttuvan, S. Sen and A. Chatterjee, ``Multidimensional adaptive power management for low-power operation of wireless devices,'' in IEEE Transactions on Circuits and Systems II: Express Briefs, vol. 55, no. 9, pp. 867-871, Sept. 2008.
		
		\bibitem{Sen2012}
		S. Sen, D. Banerjee, M. Verhelst and A. Chatterjee, ``A power-scalable channel-adaptive wireless receiver based on built-in orthogonally tunable LNA,'' in IEEE Transactions on Circuits and Systems I: Regular Papers, vol. 59, no. 5, pp. 946-957, May 2012.
		
		\bibitem{Banerjee2012}
		D. Banerjee, S. Sen, A. Banerjee, and A. Chatterjee, ``Low-power adaptive
		RF system design using real-time fuzzy noise-distortion control,'' ACM/IEEE International Symposium on Low Power Electronics and Design (ISLPED),	Jul. 2012, pp. 249-254.
		
		\bibitem{Banerjee2013}
		D. Banerjee, A. Banerjee and A. Chatterjee, ``Adaptive RF front-end design via self-discovery: using real-time data to optimize adaptation control,'' 2013 26th International Conference on VLSI Design and 2013 12th International Conference on Embedded Systems, Pune, 2013, pp. 197-202.
		
		\bibitem{Banerjee2015}
		D. Banerjee, S. K. Devarakond, X. Wang, S. Sen and A. Chatterjee, ``Real-time use-aware adaptive RF transceiver systems for energy efficiency under BER constraints,'' in IEEE Transactions on Computer-Aided Design of Integrated Circuits and Systems, vol. 34, no. 8, pp. 1209-1222, Aug. 2015.
		
		\bibitem{Banerjee2017}
		D. Banerjee, B. Muldrey, X. Wang, S. Sen and A. Chatterjee, ``Self-learning RF receiver systems: process aware real-time adaptation to channel conditions for low power operation,'' in IEEE Transactions on Circuits and Systems I: Regular Papers, vol. 64, no. 1, pp. 195-207, Jan. 2017.
		
		\bibitem{Razavi2011}
		B. Razavi, \textit{RF Microelectronics}, 2nd ed. New York, NY, USA: Prentice-Hall, 2011.
		
		\bibitem{Borremans2011}
		J. Borremans et al., ``A 40 nm CMOS 0.4–6 GHz Receiver Resilient to Out-of-Band Blockers,'' in IEEE Journal of Solid-State Circuits, vol. 46, no. 7, pp. 1659-1671, July 2011.
		
		\bibitem{Xu2018}
		J. Xu, J. Yao, L. Wang, Z. Ming, K. Wu and L. Chen, ``Narrowband Internet of Things: Evolutions, Technologies, and Open Issues,'' in IEEE Internet of Things Journal, vol. 5, no. 3, pp. 1449-1462, June 2018.
		
		\bibitem{Sen2016}
		S. Sen, ``Invited: Context-aware energy-efficient communication for IoT sensor nodes,'' 2016 53nd ACM/EDAC/IEEE Design Automation Conference (DAC), Austin, TX, 2016, pp. 1-6.
		
		\bibitem{Proakis2014}
		J. G. Proakis and M. Salehi, \textit{Digital Communications}. New York, NY, USA: McGraw Hill, 2014.		
		
		\bibitem{Goldsmith2005}
		A. Goldsmith, \textit{Wireless Communications}. New York, NY, USA: Cambridge University Press, 2005.
		
		\bibitem{Abdulaziz2016}
		M. Abdulaziz, W. Ahmad, A. Nejdel, M. T{\"o}rm{\"a}nen and H. Sj{\"o}land, ``A cellular receiver front-end with blocker sensing,'' 2016 IEEE Radio Frequency Integrated Circuits Symposium (RFIC), San Francisco, CA, 2016, pp. 238-241.
		
		\bibitem{Studer2012}
		C. Studer, S. Fateh, C. Benkeser and Q. Huang, ``Implementation trade-offs of soft-input soft-output MAP decoders for convolutional codes,'' in IEEE Transactions on Circuits and Systems I: Regular Papers, vol. 59, no. 11, pp. 2774-2783, Nov. 2012.
		
		\bibitem{Belfanti2013}
		S. Belfanti, C. Roth, M. Gautschi, C. Benkeser and Q. Huang, ``A 1Gbps LTE-advanced turbo-decoder ASIC in 65nm CMOS,'' 2013 Symposium on VLSI Circuits, Kyoto, 2013, pp. 284-285.
		
		\bibitem{Howard2006}
		S. L. Howard, C. Schlegel, and K. Iniewski, ``Error control coding in low-power wireless sensor networks: when is ECC energy-efficient?'', in EURASIP Journal on Wireless Communications and Networking, pp. 1–14, 2006.
		
		\bibitem{Grover2011}
		P. Grover, K. Woyach and A. Sahai, ``Towards a communication-theoretic understanding of system-level power consumption,'' in IEEE Journal on Selected Areas in Communications, vol. 29, no. 8, pp. 1744-1755, September 2011.
		
		\bibitem{Abramowitz1965}
		M. Abramowitz and I. Stegun, \textit{Handbook of Mathematical Functions}. New York: Dover Publications, 1970.
		
	\end{thebibliography}

\end{document}